\documentclass[12pt,a4paper]{article}
\usepackage[dvips]{graphicx}
\usepackage{amssymb}
\usepackage{epsfig}
\evensidemargin=\oddsidemargin
\renewcommand{\baselinestretch}{0.93}

\renewenvironment{thebibliography}[1]
    {\begin{list}{\arabic{enumi}.}
    {\usecounter{enumi}\setlength{\parsep}{0pt}
\setlength{\leftmargin 1.25cm}{\rightmargin 0pt}
     \setlength{\itemsep}{0pt} \settowidth
    {\labelwidth}{#1.}\sloppy}}{\end{list}}
\begin{document}
\newcommand{\lae}{\stackrel{<}{\sim}}
\newcommand{\gae}{\stackrel{>}{\sim}}
%
\def\be{\begin{equation}}
\def\ee{\end{equation}}
\def\bea{\begin{eqnarray}}
\def\eea{\end{eqnarray}}
\def\CPbar{\hbox{{\rm CP}\hskip-1.80em{/}}}
\def\D0{D\O~}
\def\pbarp{ \bar{{\rm p}} {\rm p} }
\def\pp{ {\rm p} {\rm p} }
\def\ifb{ {\rm fb}^{-1} }
\def\del{\partial }
\def\ra{\rightarrow}

\def\beq{\begin{equation}}
\def\eeq{\end{equation}}
\def\eq{\beq\eeq}
\def\beqn{\begin{eqnarray}}
\def\eeqn{\end{eqnarray}}
\relax
\def\ba{\begin{array}}
\def\ea{\end{array}}
\def\squ{\tilde{Q}}
\def\tgb{\mbox{tg$\beta$}}
\def\hH{{\cal H}}
\def\be{\begin{equation}}
\def\ee{\end{equation}}
\def\bea{\begin{eqnarray}}
\def\eea{\end{eqnarray}}
\def\pbarp{ \bar{{\rm p}} {\rm p} }
\def\pp{ {\rm p} {\rm p} }
\def\ifb{ {\rm fb}^{-1} }
\def\del{\partial }
\def\to{\rightarrow}
\def\To{\Rightarrow}
\def\tolr{\leftrightarrow}
\def\dis{\displaystyle}
\def\f{\frac}
\def\ra{\rightarrow}
\def\cbar{\bar c}
\def\bbar{\bar b}
\def\bb{b\bar b}
\def\tbar{\bar t}
\def\hbb{h b \bar b}
\def\zbb{Z b \bar b}
\def\bbbb{b \bar b b \bar b}
\def\bbjj{b \bar b j j}
\def\Psibar{\bar{\Psi}}
\def\CL{{\cal C}_L}
\def\CR{{\cal C}_R}
\def\sb{s_{\beta}}
\def\sq2{\sqrt{2}}
\def\gaga{\gamma\gamma}
\def\tanb{\tan\hspace*{-0.8mm}\beta}
\def\alphas{\alpha_s}
\def\qqbar{q\bar{q}}
\def\qqbarp{q\bar{q}'}
\def\qbarp{\bar{q}^{\prime}}
\def\tauhat{\widehat{\tau}}
\def\({\left(}
\def\){\right)}
\def\[{\left[}
\def\]{\right]}
\def\thisday{~December~1, 1998}


\def\nor{\normalsize}

\thispagestyle{empty}

\hspace{-0.90cm}  \thisday \hfill  hep-ph/9812263~\, \\ [1mm]
Physical Review D60 (1999) Oct.1  \hfill MSUHEP-81010
\vspace{1.8cm}

\renewcommand{\thefootnote}{\fnsymbol{footnote}}
\setcounter{footnote}{1}


\begin{center}
{\Large {\bf QCD Corrections to Scalar Production  
}}\\[0.14cm]
{\Large
{\bf          via Heavy Quark Fusion at Hadron Colliders}
}

\vspace{1cm}
{\sc Csaba~Bal\'azs} ~~~~ {\sc Hong-Jian He} ~~~~ {\sc C.--P.~Yuan}\\[4mm]

{\it Department of Physics and Astronomy, Michigan State University}\\[2mm]
{\it East Lansing, Michigan 48824, USA} 
\footnote{Electronic address: 
{balazs}@pa.msu.edu,~ {hjhe}@pa.msu.edu,~ {yuan}@pa.msu.edu}

\vspace{4.5cm}
\begin{abstract}
\noindent
We recently proposed that, due to the top-quark-mass enhanced Yukawa
coupling, the $s$-channel production of a charged scalar or
pseudo-scalar from heavy quark fusion can be an important new mechanism
for discovering non-standard spin-$0$ particles. In this work, we
present the complete $O(\alpha_s)$ QCD corrections to this $s$-channel
production process at hadron colliders, and also the results of QCD 
resummation over multiple soft-gluon emission. The systematic QCD-improved
production and decay rates at the FermiLab Tevatron and the CERN LHC are
given for the charged top-pions in the topcolor models, and for the
charged Higgs bosons in the generic two Higgs doublet model. The direct
extension to the production of the neutral (pseudo-)scalars via 
$b\bbar$ fusion is studied in the minimal supersymmetric standard model
(MSSM) with large $\tanb$, and in the topcolor model with large bottom
Yukawa coupling.
\\[5mm]
\noindent PACS number(s): 13.85.Ni, 12.60.Fr, 14.65.Ha, 14.80.Cp\\[0.2cm]
\noindent {\it Physical Review} D60 (1999) Oct.\,1 issue (in press)\\[0.2cm]

%
\end{abstract}
\end{center}

\baselineskip20pt   

\newpage
\renewcommand{\baselinestretch}{0.95}
\setcounter{footnote}{0}
\renewcommand{\thefootnote}{\arabic{footnote}}
\setcounter{page}{1}

\noindent
{\normalsize \bf 1.~Introduction}
\vspace*{0.3cm}

The top quark ($t$), among the three generations of fermions, is the only
one with a large mass as high as the electroweak scale. This makes
the top the most likely place to discover new physics beyond the
Standard Model (SM).
In a recent study~\cite{HY}, two of us proposed that, due to the top-mass
enhanced flavor mixing Yukawa coupling of the charm ($c$) and bottom ($b$)
with a charged scalar or pseudo-scalar ($\phi^\pm$), the $s$-channel partonic
process $c\bbar , \cbar b\to\phi^\pm$, can be an important mechanism for the
production of $\phi^\pm$ at various colliders. From the 
leading order (LO) calculation \cite{HY}, we demonstrated that the 
FermiLab Tevatron Run-II has the potential to explore the mass range of the 
charged top-pions up to about 300--350\,GeV in the 
topcolor (TopC) models \cite{TopC,TopCrev}.
In this work, we compute the complete next-to-leading order (NLO)
QCD corrections to the process $~q\qbarp\to\phi^\pm$,~
which includes the one-loop virtual corrections and the contributions
from the additional $O(\alphas )$ processes,
\be
q\qbarp\to\phi^\pm g~~~{\rm and}~~~ qg\to q'\phi^\pm .
\label{eq:alphas}
\ee
The decay width and branching ratio (BR) of such a (pseudo-)scalar
are also included up to NLO to estimate the
event rates. The QCD resummation of multiple soft-gluon radiation
is also carried out,
which provides a better prediction of the
transverse momentum distribution of the (pseudo-)scalar particle.
We shall choose the TopC model \cite{TopC} 
as a benchmark of our analysis. 
The generalization to the generic 
type-III two-Higgs doublet model (2HDM) \cite{ansatz,2HDM3}
is straightforward since the QCD-corrections are 
universal.\footnote{We note that the finite part of the counter term
to the $q$-$\qbarp$-$\phi^{0,\pm}$ Yukawa vertex is 
renormalization-scheme- and model-dependent.}
The direct extension to the production of neutral
(pseudo-)scalars via $b\bbar$ fusion is studied in  
the Minimal Supersymmetric SM (MSSM) \cite{SUSY0,SUSY} 
with large $\tanb$ and
in the TopC models with $U(1)$-tilted large bottom
Yukawa coupling \cite{TopC,TopCrev}.

\vspace{0.5cm}
\noindent
{\normalsize\bf  2.~Charged Scalar Production 
                    via Charm-Bottom Fusion }
\vspace*{0.3cm}

\noindent
{\normalsize\bf  2.1. Fixed-Order Analysis up to $O(\alpha_s)$}
\vspace{0.3cm}

We study charged (pseudo-)scalar production via the 
top-mass-enhanced flavor mixing vertex $c$-$b$-$\phi^\pm$ \cite{HY}.
The corresponding Yukawa coupling can be generally defined as
$~\CL\widehat{L}+\CR\widehat{R}~$
in which 
$~\widehat{L}=(1-\gamma_5)/2~$ and
$~\widehat{R}=(1+\gamma_5)/2$  .
The total cross sections for the $\phi^+$ production 
at hadron colliders (cf. Fig~1) can be generally expressed as
{\small
\be
\sigma \left(h_1h_2\to \phi^+X\right)=
\dis\sum_{\alpha ,\beta}
\dis\int^1_{\tau_0}dx_1\int^1_{\f{\tau_0}{x_1}}dx_2
\left[f_{\alpha/h_1}(x_1,Q^2)f_{\beta /h_2}(x_2,Q^2)
      +(\alpha\tolr\beta )\right] 
\widehat{\sigma}^{\alpha\beta}(\alpha\beta\to \phi^+X),
\label{eq:crosstot}
\ee
}
\hspace*{-1.5mm}where  
$\tau_0=m_\phi^2/S ,~
x_{1,2}\hspace*{-1mm}=\hspace*{-1.2mm}\sqrt{\tau_0}
\hspace*{1mm}\dis e^{\pm y}$,\, $m_\phi$ is the mass of $\phi^\pm$,
$\sqrt{S}$ is the center-of-mass energy of the $h_1h_2$ collider, 
and $f_{\alpha /h}(x,Q^2)$ is the 
parton distribution function (PDF) of a parton $\alpha$ with
the factorization scale $Q$.
The quantity $\widehat{\sigma}^{\alpha\beta}$ is the
partonic cross section and has the following LO contribution
for $c\bbar \to \phi^+$ 
(cf. Fig.~1a)~\cite{HY}:
\be
\dis\widehat{\sigma}^{\alpha\beta}_{\rm LO}
~=~\dis\delta_{\alpha c}\delta_{\beta\bbar}
\delta (1-\widehat{\tau})\widehat{\sigma}_0~,~~~~~
\widehat{\sigma}_0 \equiv
\f{\pi}{12\widehat{s}}\left(|\CL|^2+|\CR|^2\right),
\label{eq:sigmaLO}
\ee
where ~$\widehat{\tau}=m_\phi^2/\widehat{s}$~ with  $\widehat{s}$ the
center-of-mass energy of the sub-process, and 
the terms suppressed by the small mass ratio
$(m_{c,b}/m_\phi )^2$ have been ignored.
Since we are interested in the inclusive production of the scalar
$\phi$, it is natural to choose the factorization scale $Q$ to be its
mass $m_\phi$, which is of $O(10^{2-3})$\,GeV and much larger than
the mass of charm or bottom quark. Hence, in this work, we will treat
$c$ and $b$ as massless partons inside proton or antiproton and perform 
a NLO QCD calculation with consistent sets of 
PDFs \cite{Haber,ACOT,Collins}.

The NLO contributions are of $O(\alphas )$, which contain three
parts: (i)~the one-loop Yukawa vertex and quark self-energy corrections
(cf. Fig.~1b-d); (ii)~the real gluon emission in the $q\qbarp$-annihilations
(cf. Fig.~1e); (iii)~$s$- and $t$-channel gluon-quark fusions
(cf. Fig.~1f-g). The Feynman diagrams coming from permutations 
are not shown in Fig.~1. 
Unlike the usual Drell-Yan type of processes (where the
sum of the one-loop quark-wavefunction renormalization and vertex correction
gives the ultraviolet finite result), we need to include
the renormalization for the Yukawa coupling ($y_j$) which usually 
relates to the relevant quark mass ($m_{q_j}$), 
i.e., we have to add the counter term  
at the NLO (cf. Fig.1d)
besides the contribution from the usual wavefunction renormalization 
$~Z_{q_1q_2\phi}=\f{1}{2}(Z_{q_1}+Z_{q_2})$  (cf. Fig.1c).
This applies to the
Yukawa interactions of the SM and MSSM Higgs bosons 
as well as the top-pions in the TopC models. 
It is clear that, for flavor-mixing vertex 
$c$-$b$-$\phi^\pm$ in the TopC model [cf. Eq.~(\ref{eq:Ltoppi}) below],   
the counter-term of the Yukawa
coupling is equal to the top quark mass counter-term $\delta m_t/m_t$, 
which we determine from the top-quark mass renormalization
in the on-shell scheme so that $m_t$ is the pole mass of the top quark. 
In other cases such as in the general 2HDM (type-III) \cite{2HDM3} 
and the TopC models (with $b$-Higgs or $b$-pions) \cite{TopC2}, 
some of their Yukawa couplings are not related to 
quark masses or not of the above simple one-to-one correspondence, 
and thus have their independent counter terms ($\delta y_j/y_j$).
In addition to the virtual QCD-loop corrections, 
the contributions of the real gluon emission
from the initial state quarks have to be included (cf. Fig.~1e).
The soft and collinear singularities appeared in these diagrams
are regularized by the dimensional regularization prescription
at $D=4-2\epsilon$ dimensions. After summing up the contributions of
virtual gluon-loop and real gluon-radiation (cf. Fig.~1b-e),  the 
ultraviolet and soft singularities separately cancel. 
The collinear singularities
are still left over and should be absorbed into the renormalization
of the PDF \cite{PDF-ren}. (The $\overline{\rm MS}$ renormalization scheme
is used in our calculation.)
Finally, the gluon-quark fusion sub-processes 
(cf. Fig.~1f-g) should also be taken into account and computed at general
dimension-$D$. All these results are separately
summarized into the Appendix.  

\begin{figure}[ht]
\begin{center}
\includegraphics[width=16cm,height=12cm]{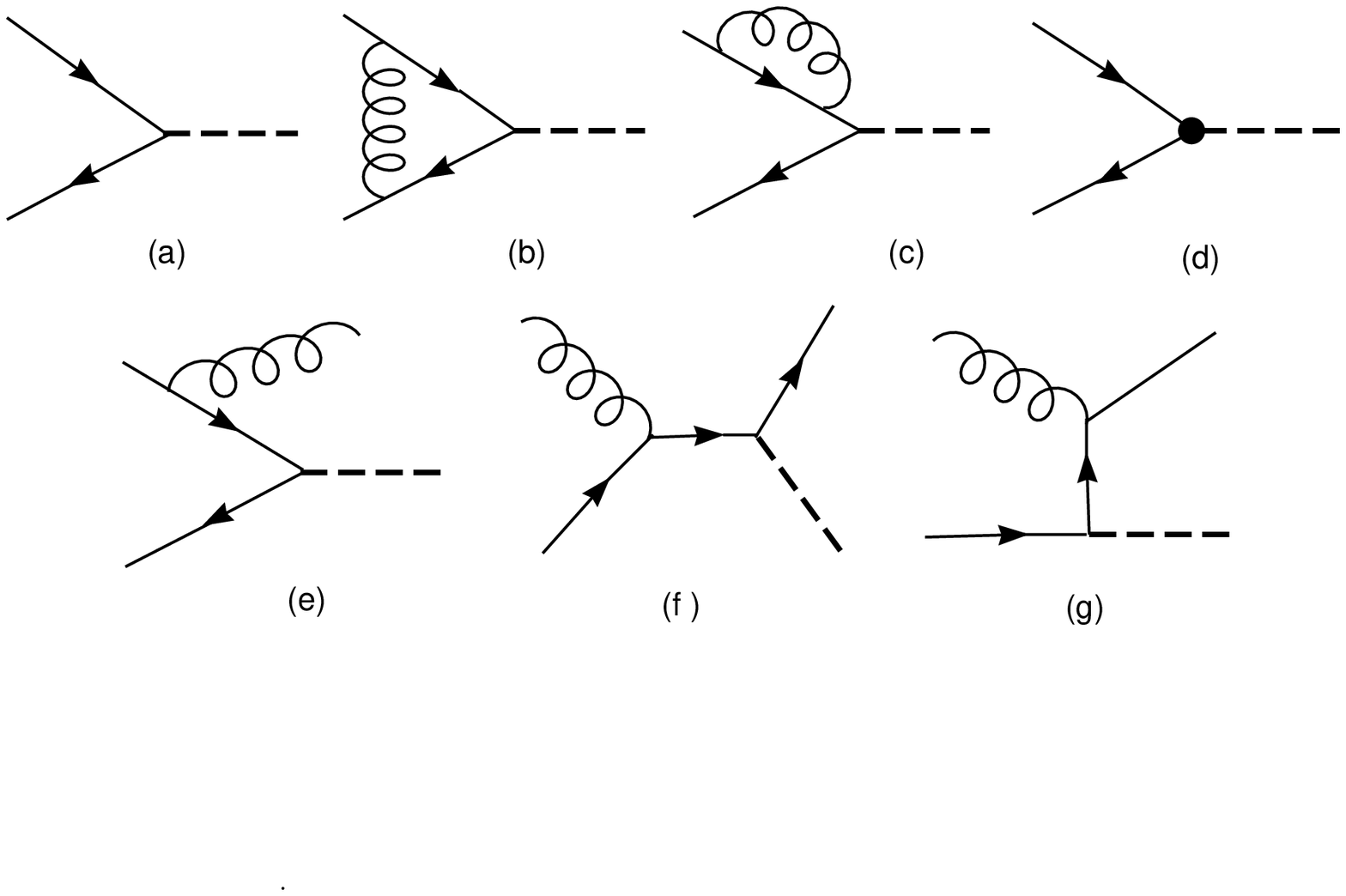}
\end{center}
\vspace{-5.3cm}
\caption{\small 
Representative diagrams for charged or neutral (pseudo-)scalar
(dashed line) production from quark-antiquark and quark-gluon collisons at 
$O(\alphas^0)$ and $O(\alphas^1)$:
(a)~leading order contribution; 
(b-d)~self-energy and vertex corrections (with counter term);
(e)~real gluon radiation in $q\qbarp$-fusion;
(f-g)~$s$- and $t$-channel gluon-quark fusions.
}
\label{fig:graphs}
\end{figure}

The hadron cross sections become regular
after renormalizing the Yukawa coupling and
the PDFs in (\ref{eq:crosstot}), which 
are functions of the renormalization scale $\mu$ and the
factorization scale $\mu_F(=\sqrt{Q^2})$. 
The partonic NLO cross section $\widehat{\sigma}^{\alpha\beta}_{\rm NLO}
\(\alpha\beta\to\phi^+X\)$ contains the contributions
$\Delta\widehat{\sigma}_{q\qbarp}(q\qbarp\to\phi^+,\,\phi^+g)$, 
$\Delta\widehat{\sigma}_{qg}(qg\to\phi^+q' )$, and 
$\Delta\widehat{\sigma}_{\bar{q}g}(\bar{q}g\to\phi^+\qbarp)$: 
\be
\ba{l}
  \(\Delta\widehat{\sigma}_{q\qbarp},~
    \Delta\widehat{\sigma}_{qg},~
    \Delta\widehat{\sigma}_{\bar{q}g}\)
=\dis\widehat{\sigma}_0\times\f{\alphas}{2\pi}
\left( \delta_{qc}\delta_{\qbarp \bbar}
\Delta\overline{\sigma}_{c\bbar},~
\delta_{qc}\Delta\overline{\sigma}_{cg},~
\delta_{\bar{q}\bbar}\Delta\overline{\sigma}_{\bbar g}
\right),\\[4mm]
\Delta\overline{\sigma}_{c\bbar}
=\dis C_F
\left[
 4\left(1+\tauhat^2\right)\left(\f{\ln (1-\tauhat )}{1-\tauhat}\right)_+
-2\f{1+\tauhat^2}{1-\tauhat}\ln\tauhat
+\left(\f{2\pi^2}{3}-2-\Omega\right)\delta (1-\tauhat )
+2(1-\tauhat )      \right]\\[4mm]
~~~~~~~~~+\dis 2P^{(1)}_{q\leftarrow q}(\tauhat )\ln\f{m_\phi^2}{Q^2}~,  
\\[4mm]
\Delta\overline{\sigma}_{cg,\bbar g}=
\dis P^{(1)}_{q\leftarrow g}(\tauhat )
\[\ln\f{\(1-\tauhat\)^2}{\tauhat} +\ln\f{m_\phi^2}{Q^2}\]
-\f{1}{4}(1-\tauhat )\(3-7\tauhat\), \\[6mm]
\dis P^{(1)}_{q\leftarrow q}(\tauhat )=
\dis C_F\(\f{1+\tauhat^2}{1-\tauhat}\)_+, ~~~~
\dis P^{(1)}_{q\leftarrow g}(\tauhat )=
\f{1}{2}\[\tauhat^2+(1-\tauhat )^2\],
\ea
\label{eq:NLO}
\ee
where  $~\tauhat =m_\phi^2/\widehat{s}$ and $C_F=4/3$.
The mass counter term for the Yukawa vertex renormalization 
is determined in the on-shell scheme, i.e.,
\be
\dis\f{\delta m_t}{m_t}
=-\f{C_F\alpha_s}{4\pi}\[3\(\f{1}{\epsilon}-\gamma_E+\ln 4\pi \)
+\Omega \],
\label{eq:delta_mt}
\ee
in the TopC model.
Here, the bare mass $m_{t0}$ and the renormalized mass $m_{t}$ 
are related by $\,m_{t0}=m_t+\delta m_t$\, and  
$m_t\simeq 175$\,GeV is taken to be the top-quark pole mass. 
The finite part of the mass counter term is
$~\dis\Omega = 3\ln \left[\mu^2/m_t^2\right]+4~$ in the TopC model, 
where $\Omega \geq 0$ for $\mu \geq m_te^{-2/3}\simeq 90$~GeV.
In the following, we shall choose the QCD factorization scale $\mu_F$ 
(set as the invariant mass ${\small \sqrt{Q^2}}$) and the 
renormalization scale $\mu$ to be the same as the scalar mass,
i.e.,   ${\small \sqrt{Q^2}}=\mu=m_\phi$, 
which means that in (\ref{eq:NLO})  
the factor $\ln\(m_\phi^2/Q^2\)$ vanishes and 
the quantity $\Omega$ becomes 
\be
\dis\Omega = 3\ln \left[m_\phi^2/m_t^2\right]+4~.
\label{eq:omega_final}
\ee

For the case of $m_\phi\gg m_t$, 
the logarithmic term $\ln \(m_\phi^2/m_t^2\)$
becomes larger for $m_\phi \gg m_t$, and its contributions to all orders
in $\alphas\ln\(m_\phi^2/m_t^2\)$ may be resummed by introducing
the running Yukawa coupling $y_t(\mu )$, or correspondingly, the
running mass $m_t(\mu )$. 
In the above formula, $m_t$ is the pole mass ($m_t^{\rm pol}\simeq 175$\,GeV)
and is related to the one-loop running mass via the relation \cite{mtrun}
\be
\dis m_t(\mu )=m_t( m_t^{\rm pol})
\[1-\f{3C_F}{4\pi}\alpha_s(\mu )\ln\f{\mu^2}{m_t^{\rm pol}}\],~~~~~
m_t( m_t^{\rm pol})
=m_t^{\rm pol}\[1+\f{C_F}{\pi}\alpha_s(m_t^{\rm pol})\]^{-1} .
\label{eq:mtrun1loop}
\ee
Using the renormalization group equation, one can resum the leading
logarithms to all orders in $\alphas$ \cite{peskin} and obtains 
\be
\dis m_t(\mu )=m_t( m_t^{\rm pol})
\[\f{\alpha_s (\mu )}{\alpha_s(m_t^{\rm pol})}\]^{\f{9C_F}{33-2n_f}} ,
\label{eq:mtrunsum}
\ee
with $n_f=6$ for $\mu >m_t$ .
Thus, to include the running effect of the Yukawa coupling,  
we can replace the $(m_t^{\rm pol})^2$-factor (from the Yukawa
coupling) inside the square of the $S$-matrix element
[up to $O(\alpha_s)$] by the running factor
\be
\dis m_t^2(\mu )\left\{1+2\f{C_F\alpha_s(\mu )}{\pi}\[1+\f{3}{4}
\ln\(\f{\mu}{m_t^{\rm pol}}\)^2 \] \right\}
\,=\, m_t^2(\mu ) \[1+\f{C_F\alpha_s(\mu )}{2\pi}\Omega\] ,
\label{eq:runfactor}
\ee
where the logarithmic term in the bracket $[\cdots ]$ 
is added to avoid double-counting with the resummed logarithms 
inside $m_t^2(\mu )$.  It is clear
that this  $~\dis\[1+\(C_F\alpha_s(\mu )/2\pi\)\Omega\]~$ 
factor will cancel the $\Omega$-term inside the NLO hard cross section 
$\Delta\widehat{\sigma}_{c\bbar}$ in Eq.~(\ref{eq:NLO}) at $O(\alphas )$, 
so that the net effect of the Yukawa vertex renormalization
(after the resummation of leading logarithms) is to replace the relevant 
tree-level on-shell quark mass (related to the Yukawa coupling) by its 
$\overline{\rm  MS}$ running mass [cf. Eq.~(\ref{eq:mtrunsum})] and  remove
the $\Omega$-term in Eq.~(\ref{eq:NLO}).
When the physical scale $\mu$ (chosen as the scalar mass 
$m_\phi$) is not much larger than $m_t$, 
the above running effect is small since the $\ln (\mu/m_t)$ factor in
the Yukawa counter-term $\delta m_t/m_t$ is small.  
However, the case 
for the neutral scalar production via the $b\bbar$ annihilation can be
different. When the loop correction to the $\phi^0$-$b$-$\bbar$ 
Yukawa coupling contains the logarithm $\ln (\mu /m_b)$, which is much
larger than $\ln (\mu /m_t)$,
these large logarithms should be resummed into the running coupling,
as we will do in Section~4.

In the TopC model, there are three pseudo-scalars, called top-pions,
which are predicted to be light, with a mass of $O(100 \sim 300)$
GeV. The relevant Yukawa interactions for top-pions,
including the large $t_R$-$c_R$ flavor-mixing, can be written 
as\footnote{As pointed out in Ref.~\cite{HY},
an important feature deduced from (\ref{eq:Ltoppi}) 
is that the charged top-pion 
$\pi_t^\pm$ mainly couples to the right-handed top 
($t_R$) or charm ($c_R$) but not
the left-handed top ($t_L$) or charm ($c_L$), 
in contrast to the standard $W$-$t$-$b$
coupling which involves only $t_L$. This makes the top-polarization
measurement very useful for further discriminating the signal from the 
background events.}
\cite{HY}
\be
\ba{ll}
{\cal L}_Y^{\pi_t}~=&
-\dis\f{m_t\tanb}{v}\left[
iK_{UR}^{tt}{K_{UL}^{tt}}^{\hspace*{-1.3mm}\ast}\overline{t_L}t_R\pi_t^0
+\sq2
{K_{UR}^{tt}}^{\hspace*{-1.3mm}\ast}K_{DL}^{bb}\overline{t_R}b_L\pi_t^+ 
+ \right. \\[3mm]
&~~~~~~~~~~~~~~\left.
iK_{UR}^{tc}{K_{UL}^{tt}}^{\hspace*{-1.3mm}\ast}\overline{t_L}c_R\pi_t^0
+\sq2
{K_{UR}^{tc}}^{\hspace*{-1.3mm}\ast}K_{DL}^{bb}\overline{c_R}b_L\pi_t^+ 
+{\rm h.c.}       \right],
\ea
\label{eq:Ltoppi}
\ee
where {\small $\tanb = \sqrt{(v/v_t)^2-1}$}$\,\sim\! O(4\! -\! 1.3)$ 
with the top-pion decay constant $v_t\!\sim\! O(60\!-\!150)$\,GeV, 
and the full vacuum expectation value (VEV) 
$v\simeq 246$~GeV (determined by the Fermi constant).
The analysis from top-quark decay in the Tevatron $t\bar t$ events
sets a direct lower bound on the charged top-pion mass 
to be larger than about 150\,GeV \cite{bound,TopC}. 
The existing low energy LEP/SLD measurement of $R_b$, 
which slightly lies above the SM value 
by about 0.9$\sigma$ \cite{hollik}, also provides an indirect 
constraint on the top-pion Yukawa coupling 
${\cal C}_R^{tb}=(\sqrt{2}m_t/v)\tanb$ due to the one-loop contribution
of charged top-pions to $R_b$. 
However, given the crude approximation in estimating the top-pion
loops (with all higher order terms ignored)
and the existence of many other sources of 
contributions associated with the strong dynamics, the indirect
$R_b$ constraint is not conclusive \cite{TopC}.
For instance, it was shown that the 3$\sigma$ $R_b$ bound from 
the one-loop top-pion correction can be fully
removed if the top-pion decay
constant $v_t$ is increased by about a factor of 2
(which is the typical uncertainty of the Pagels-Stokar 
estimate)\,\cite{TopC,Rb}; also, the non-perturbative contributions of 
the coloron-exchanges can shift the $R_b$ above its SM 
value\,\cite{TopC} and tend to cancel the negative top-pion corrections.
Due to these reasons, it is clear that the inconlusive $R_b$-bound in 
the TopC models should not be taken too seriously. 
Nevertheless, to be on the safe side, we will 
impose the roughly estimated $R_b$-constraint in our current analysis of
the TopC model, by including {\it only}
the (negative) one-loop top-pion contribution as in 
Ref.\,\cite{Rb}.\footnote{
However, it is important to keep in mind 
that such a rough $R_b$-bound is likely to
over-constrain the top-pion Yukawa coupling since only the negative 
one-loop top-pion correction (but nothing else) is included in this 
estimate. A weaker $R_b$-bound will less reduce the top-pion Yukawa
coupling and thus allow larger production rates of charged top-pions at
colliders which can be obtained from our current analysis by simple
re-scaling.}~
As shown in Fig.\,\ref{fig:Rb}a, the current 3$\sigma$ $R_b$-bound
requires a smaller top-pion Yukawa coupling,
${\cal C}_R^{tb}\!\sim\! 1.3-2$ (or, $\tanb \!\sim\! 1.3-2$), 
for the low mass region of $m_{\pi_t^\pm}\!\sim\! 200-500$\,GeV.
Since the top-pion decay constant $v_t$ is related to $\tanb$,
this also requires $v_t$ to be around 
$150\!\sim\!100$\,GeV for $m_{\pi_t^\pm}\!\sim\! 200-500$\,GeV
(cf. Fig.\,\ref{fig:Rb}b). For comparison, 
the usual Pagels-Stokar estimate of $v_t$ (by keeping only the
leading logarithm but not constant terms),
$v_t^2=(N_c/8\pi^2)m_t^2\ln\Lambda^2/m_t^2$,
gives $v_t\!\sim\! 64\!-\!97$\,GeV for the topcolor breaking scale 
$\Lambda\!\sim\! 1\!-\!10$\,TeV, where a typical factor of $2\!\sim\!3$
uncertainty in the calculation of $v_t^2$
is expected \,\cite{TopC,PS}.
This estimate is slightly lower than the $R_b$-constrained values of $v_t$ 
in Fig.\,\ref{fig:Rb}b, but is still in reasonable consistency
(given the typical factor of $\sqrt{2}\!\sim\!\sqrt{3}$
error in the leading logarithmic Pagels-Stokar estimate of $v_t$).

In (\ref{eq:Ltoppi}), $K_{UL,R}$ and $K_{DL,R}$ are defined from 
diagonalizing the up- and down-type quark mass matrices $M_U$ and $M_D$:
{\small ~$K_{UL}^\dag M_U K_{UR} = M_U^{\rm dia},~
K_{DL}^\dag M_D K_{DR} = M_D^{\rm dia}~,$}
with {\small $M_U^{\rm dia}$}$={\rm diag}(m_u,m_c,m_t)$ and
     {\small $M_D^{\rm dia}$}$={\rm diag}(m_d,m_s,m_b)$.
For the class-I TopC models \cite{TopC2}, 
we have constructed \cite{HY}  a realistic and attractive
pattern of $K_{UL}$ and $K_{DL}$ so that the well-constrained 
Cabibbo-Kobayashi-Maskawa (CKM) matrix
$V~(=K_{UL}^\dag K_{DL})$ can be reproduced 
in the Wolfenstein-parametrization
\cite{Wolfen} and all potentially large contributions 
to the low energy data (such as the
$K$-$\bar K$, $D$-$\bar D$ and $B$-$\bar B$ mixings 
and the $b\to s\gamma$ rate) can be avoided~\cite{HY}. 
We then found that the right-handed rotation matrix $K_{UR}$ is 
constrained such that its 33 and 32 elements take the values as~\cite{HY}
\be          
K_{UR}^{tt}\simeq 0.99\hspace*{-0.7mm}-\hspace*{-0.7mm}0.94,~~~
K_{UR}^{tc}\leq \sqrt{1-{K_{UR}^{tt}}^2} 
\simeq 0.11\hspace*{-0.7mm}-\hspace*{-0.7mm}0.33,
\label{eq:KURtc}
\ee
which show that the $t_R$-$c_R$ flavor mixing 
can be naturally around $10-30\%$.

For the current numerical analysis 
we consider a benchmark choice \cite{HY} 
based upon the above TopC model:
\be
{\cal C}^{tb}_R= {\cal C}_R^{tb}(R_b{\rm \,constrained}), ~~~~
{\cal C}^{cb}_R={\cal C}^{tb}_R{K_{UR}^{tc}}
               \simeq {\cal C}^{tb}_R\times 0.2,~~~~
{\cal C}_L^{tb}={\cal C}_L^{cb}=0.
\label{eq:Coupling0}
\ee
It is trivial to scale the numerical results presented in this paper to
any other values of ${\cal C}_{L,R}$ when needed. Unless specified otherwise, 
we use CTEQ4M PDF \cite{CTEQ4M} to calculate the rates. Note that
CTEQ4M PDFs are consistent with the scheme used in the current study
which treats the initial state quarks as massless partons in computing
the Wilson coefficient functions. The only effect of the heavy quark mass is
to determine at which scale $Q$ this heavy quark parton becomes 
active.\footnote{This is the Collins-Wilczek-Zee (CWZ) scheme
\cite{CWZ}.} In our case, the scale $Q=m_\phi\gg m_c,\,m_b$.

In Fig.~\ref{fig:FigSigma_FixMt.eps}, 
we present the total cross sections for
the charged top-pion production as functions of its mass, 
at the Tevatron 
(a $p\bar p$ collider at 1.8 and 2\,TeV) and 
the LHC (a $pp$ collider at 14\,TeV).
We compare the improvements by including the complete NLO results
[cf. (\ref{eq:NLO})] and by including the resummed running Yukawa coupling
or running mass [cf. (\ref{eq:mtrunsum})].
For this purpose, we first plot the LO total cross sections with the 
tree-level Yukawa coupling [dash-dotted curves, 
cf. (\ref{eq:sigmaLO}) and (\ref{eq:Coupling0})] and 
with the resummed running Yukawa coupling 
or running mass [dotted curves, cf. (\ref{eq:sigmaLO}) and (\ref{eq:mtrunsum})]; 
then we plot the NLO cross sections with the 
one-loop Yukawa coupling [dashed curves, cf.
(\ref{eq:NLO})] and with the resummed running Yukawa coupling or running mass 
[solid curves, cf. 
(\ref{eq:NLO}), (\ref{eq:mtrunsum}) and (\ref{eq:runfactor})] .
We see that at the LHC there is a visible difference between the
pure LO results with tree-level Yukawa coupling (dash-dotted curves) and
other NLO and/or running-coupling improved results. But at the Tevatron,
the LO results with running Yukawa coupling (dotted curves) are visibly
smaller than the results in all other cases for $m_\phi >300$\,GeV. 
This shows that without the complete NLO calculation, including only the 
running Yukawa coupling in a LO result may not always warrant a 
better improvement. 
Finally, the comparison in Fig.~\ref{fig:FigSigma_FixMt.eps}
shows that the resummed running Yukawa coupling or top-mass
[cf. Eq.~(\ref{eq:mtrunsum})]
does not generate any significant improvement from the one-loop running.
This is because the top-mass is large and
$\alphas\ln\(m_\phi^2/m_t^2\)$ is small for $m_\phi$ up to 1\,TeV. Thus, the
improvement of the resummation in (\ref{eq:mtrunsum}) has to come from
higher order effects of $\alphas\ln\(m_\phi^2/m_t^2\)$. However, as to be
shown in Sec.~4, the situation for summing over powers of 
$\alphas\ln\(m_\phi^2/m_b^2\)$  is different due to 
$m_b\ll m_t,\, m_\phi$\,. 

Fig.~\ref{fig:Sigma1} is to examine the individual NLO contributions 
to the charged top-pion production via the $\qqbar '$ and $qg$
sub-processes,
in comparison with the full NLO contributions.\footnote{
Unless specified, $qg$ includes both $qg$ and ${\bar q}g$ contributions.}
The LO contributions are also shown as a reference.\footnote{
With the exception of Figs.~\ref{fig:FigSigma_FixMt.eps}, ~\ref{fig:GammaBr},
and ~\ref{fig:Sigma_MSSMbb}, we only show our numerical results with the 
resummed running Yukawa coupling or running mass.} 
[Here $q$ denotes the heavy charm or bottom quark.] 
In this figure, there are
three sets of curves for the charged top-pion production
cross sections: the highest set 
is for the LHC ($\sqrt{S}=14$\,TeV), the middle set is 
for the upgraded Tevatron ($\sqrt{S}=2$\,TeV),
and the lowest set is for the Tevatron Run~I ($\sqrt{S}=1.8$\,TeV).
The LO cross sections are plotted as dotted lines while the NLO cross sections
as solid ones. The dashed lines show the contributions from the
$q\bar{q}'$-fusion sub-processes, 
and the dash-dotted lines describe the contributions
from the $qg$-fusion sub-processes. 
The $qg$-fusion cross sections are negative
and are plotted by multiplying a factor of $-1$, for convenience. 
For a quantitative comparison 
of the individual NLO contributions versus the full NLO
results, we further plot, in  Fig.~\ref{fig:InitKFac}, 
the ratios (called $K$-factors) of the different NLO contributions 
to the LO cross section by using the same set of CTEQ4M PDFs.
The solid lines of Fig. \ref{fig:InitKFac} show
that the overall NLO corrections to the $pp,p\bar{p} \to \phi^\pm X$
processes are positive for $m_\phi$ above $\sim$150\,(200)~GeV
and lie below $\sim$15\,(10)\% for the
Tevatron (LHC) in the relevant mass region. This is in contrast with
the NLO corrections to the $W^\pm$ boson production at hadron
colliders, which are always positive and as large as about
25\% at the Tevatron \cite{Balazs-Yuan-PRD97}.   
The reason
of this difference originates from the differences in the $\Delta \sigma_ 
{q\bar{q}'}$ and $\Delta \sigma _{qg,g\bar{q}}$ for $\phi^\pm$ and $W^\pm$
production. While in the case of $W^\pm$ production the positive 
$\Delta \sigma_{q\bar{q}'}$ piece dominates, in the case of  $\phi^\pm$
production the size of negative $\Delta \sigma _{qg,g\bar{q}}$ piece becomes
comparable with that of the positive $\Delta \sigma_{q\bar{q}'}$\,
such that a non-trivial cancellation occurs.

While it is reasonable to take the renormalization and the factorization
scales to be $m_\phi$ for predicting the inclusive production rate of
$\phi^+$, it is desirable to estimate the uncertainty in the rates due to
different choices of PDFs. For that purpose, we examine a few typical
sets of PDFs from CTEQ4, which predict different shapes of charm, bottom
and gluon distributions. As shown in Table~\ref{tb:PDFS} and 
Fig.~\ref{fig:PDF}, the uncertainties due to the choice of PDF set are 
generally within $\pm 20\%$ for the relevant scalar mass ranges 
at both the Tevatron and the LHC.
\begin{table}[ht]
\vspace*{-5mm}
\caption{\small 
Cross sections in fb for charged top-pion production in the TopC model at
the upgraded Tevatron and the LHC are shown, by using four different CTEQ4 
PDFs. They are separately given for the LO and NLO processes, 
and for the $q\bar{q}\to\phi^+X$ and $qg\to\phi^+X$ sub-processes. 
At the upgraded Tevatron the top number is for $m_{\phi} = 200$ GeV, 
the middle is for $m_{\phi} = 300$ GeV, and
the bottom is for $m_{\phi} = 400$ GeV.
At the LHC the top number is for $m_{\phi} = 400$ GeV, 
the middle is for $m_{\phi} = 700$ GeV, and
the lowest is for $m_{\phi} = 1$ TeV.
}
\begin{center}
\begin{tabular}{c||r r r r|r r r r}
\hline\hline
&&&&&&&&\\[-0.2cm]
Collider & 
\multicolumn{4}{c|}{Upgraded Tevatron (2\,TeV)} & 
\multicolumn{4}{c}{LHC (14\,TeV)}  \\
[0.15cm] \cline{1-9}
&&&&&&&&\\[-0.2cm]
Process $\backslash$ PDF 
                    &   4A1 &    4M &   4A5 &   4HJ
                    &   4A1 &    4M &   4A5 &   4HJ \\
[0.15cm]\hline\hline
&&&&&&&&\\[-0.2cm]
                    &   367 &   382 &   376 &   387
                    &  5380 &  5800 &  6060 & 5890 \\
&&&&&&&&\\[-0.3cm]
LO                  &  42.6 &  43.7 &  41.5 &  46.6
                    &   863 &   901 &   896 &   906 \\
&&&&&&&&\\[-0.3cm]
                    &  6.88 &  7.05 &  6.56 &  8.38
                    &   235 &   240 &   232 &   241 \\
[0.2cm]\hline
&&&&&&&&\\[-0.2cm] 
                    &   370 &   402 &   412 &   407
                    &  5430 &  6080 &  6510 &  6170 \\
&&&&&&&&\\[-0.3cm]
NLO                 &  45.6 &  48.6 &  47.9 &  51.6
                    &   912 &   976 &   997 &   981 \\
&&&&&&&&\\[-0.3cm]
                    &  7.70 &  8.21 &  7.89 &  9.56
                    &   255 &   266 &   264 &   268 \\
[0.2cm]\hline
&&&&&&&&\\[-0.2cm] 
                    &   551 &   584 &   585 &   590
                    &  7530 &  8290 &  8740 &  8400 \\
&&&&&&&&\\[-0.3cm]
$q\bar{q}\to\phi^+X$&  64.5 &  67.4 &  65.5 &  71.7
                    &  1210 &  1280 &  1290 &  1290 \\
&&&&&&&&\\[-0.3cm]
                    &  10.6 &  11.1 &  10.5 &  13.0
                    &   331 &   341 &   335 &   343 \\
[0.2cm]\hline
&&&&&&&&\\[-0.2cm] 
                    &$-$ 180&$-$ 181&$-$ 174&$-$ 183
                    &$-$2100&$-$2200&$-$2240&$-$2230\\
&&&&&&&&\\[-0.3cm]
$qg\to\phi^+X$      &$-$19.2&$-$18.9&$-$17.5&$-$19.9
                    & $-$299& $-$302& $-$293& $-$303\\
&&&&&&&&\\[-0.3cm]
                    &$-$2.94&$-$2.86&$-$2.59&$-$3.34
                    &$-$76.0&$-$74.7&$-$70.6&$-$75.0\\
[0.2cm]\hline\hline 
\end{tabular}
\end{center}
\label{tb:PDFS}
\end{table}

\vspace{0.4cm}
\noindent
{\normalsize\bf  2.2. Analysis of Multiple Soft-Gluon Resummation}
\vspace{0.3cm}

The $\alpha _s$ corrections to the (pseudo-)scalar production
involve the contributions from the emission of virtual and real
gluons, as shown in Figs.~1(b), (c) and (e). As the result of the real gluon
radiation, the (pseudo-)scalar particle will acquire a non-vanishing
transverse momentum ($Q_T$). When the emitted gluons are soft, they generate
large logarithmic contributions of the form (in the lowest order): 
$\alpha _s\ln ^m\left(Q^2/Q_T^2\right) /Q_T^2$, 
where $Q$ is the invariant mass of the
(pseudo-)scalar, and $m=0,1$. These large logarithms spoil the convergence of
the perturbative series, and falsify the $O(\alpha _s) 
$ prediction of the transverse momentum when $Q_T\ll Q$.

To predict the transverse momentum distribution of the produced 
(pseudo-)scalar, we utilize the Collins--Soper--Sterman (CSS) 
formalism \cite{CSS}, 
resumming the logarithms of the type \linebreak
$\alpha _s^n \ln ^m\left( Q^2/Q_T^2\right)
/Q_T^2$~, to all orders $n$ in $\alpha _s$ ($m=0,...,2n-1$). 
The resummation calculation is
performed along the same line as for vector boson production in
Ref.~\cite{Balazs-Yuan-PRD97}. Here we only give the differences from 
that given in Ref.~\cite{Balazs-Yuan-PRD97}. But for convenience, 
we also list the $A^{(1)}$, $A^{(2)}$, and $B^{(1)}$ coefficients of the 
Sudakov exponent, which have been used in the current analysis:
\be
\ba{l}
A^{(1)}\(C_1\) \,=\, C_F\,,~~~~~\,
\dis B^{(1)}\(C_1 =b_0,C_2=1\)\,=\,-\frac{3}{2}C_F\,,  \\[3mm]
\dis A^{(2)}\(C_1 =b_0\)\,=\,
C_F\left[ \left( {\frac{67}{36}}-{\frac{\pi ^2}{12}}
\right) N_C-\f{5}{18}n_f\right],
\ea
\label{eq:A12B1}
\ee
where $C_F=4/3$ is the Casimir of the fundamental representation of SU(3), 
$N_C=3$ is the number of SU(3) colors, and $n_f$ is the
number of light quark flavors with masses less than $Q$. 
In the above we used the
canonical values of the renormalization constants $C_1=b_0$, and $\ C_2=1$. 

To recover the $O\left( \alpha _s\right) $
total cross section, we also include the Wilson coefficients $C_{i\alpha
}^{(1)}$, among which $C_{ij}^{(1)}$ differs from the vector boson
production (here $i$ denotes quark or antiquark flavors, and 
$\alpha = q_i$ or gluon $g$). Explicitly,
\begin{eqnarray}
C_{jk}^{(0)}(z,b,\mu ,{{C_1}/{C_2}}) & \hspace*{-8mm}
=~\dis\delta _{jk}\delta ({1-z}), 
~~~~
C_{jg}^{(0)}(z,b,\mu ,{C_1}/{C_2}) ~=~0, 
\nonumber \\[2mm]
C_{jk}^{(1)}(z,b,\mu ,{{C_1}/{C_2}}) &
\hspace*{-8mm}=~ \dis\delta _{jk}C_F\left\{ \frac
12(1-z)-\frac 1{C_F}\ln \left( \frac{\mu b}{b_0}\right) P_{j\leftarrow
k}^{(1)}(z)\right.  
\nonumber \\[3mm]
& ~~~~~\dis\left. +\,\delta (1-z)\left[ -\ln ^2\left( {\frac{C_1}{{b_0C_2}}}
e^{-3/4}\right) +{\frac{\mathcal{V}}4}+{\frac 9{16}}\right] \right\}, 
\nonumber \\[3mm]
C_{jg}^{(1)}(z,b,\mu ,{{C_1}/{C_2}}) &
\hspace*{-27mm}=~\dis\f{1}{2}z(1-z)-\ln \left( 
\frac{\mu b}{b_0}\right) P_{j\leftarrow g}^{(1)}(z),
\label{eq:C01}
\end{eqnarray}
where $P_{j\leftarrow g}^{(1)}$ is the $O(\alphas )$ gluon splitting kernels
\cite{DGL,AP} given in the Appendix.
In the above expressions, $\mathcal{V=V}_{DY}=-8+\pi ^2$ for the vector boson
production~\cite{Balazs-Yuan-PRD97},  and 
$\mathcal{V=V}_\Phi =\pi ^2$ for the (pseudo-)scalar
production, when using the running mass given in Eq.~(\ref{eq:mtrunsum}) 
for the Yukawa coupling. Using the canonical values of the
renormalization constants, $\ln (\mu b/b_0)$ vanishes, because 
$\mu=C_1/b=b_0/b$.

The only remaining difference between the resummed formulae of the vector
boson and (pseudo-)scalar production is in the regular ($Y$) terms, which
comes from the difference of the  $O\left( \alpha_s\right) $ 
real emission amplitude squares (cf., the definitions of
${\cal T}_{q\overline{q}}^{-1}$ and ${\cal T}_{qg}^{-1}$ 
in Appendix C of Ref.~\cite{Balazs-Yuan-PRD97} and 
Eqs.~(\ref{eq:M2qq}) and (\ref{eq:M2gq}) of this paper). 
The non-perturbative sector of the CSS resummation 
(the non-perturbative function and the related parameters) 
is assumed to be the same as that in Ref.~\cite{Balazs-Yuan-PRD97}.

As described in Ref.~\cite{Balazs-Yuan-PRD97}, the resummed total rate
is the same as the $O(\alpha _s)$ rate, when we include 
$C^{(1)}_{i\alpha}$ and $Y^{(1)}$, 
and switch from the resummed distribution to the fixed order one
at $Q_T=Q$. When calculating the total rate, we have applied this matching
prescription. In the case of the (pseudo-)scalar production, 
the matching takes place at high $Q_T \sim Q$ values, 
and the above matching prescription is irrelevant 
when calculating the total rate
because the cross sections there are negligible. 
Thus, as expected, the resummed total rate differs from the 
$O(\alpha _s)$ rate only by a few percent. 
Since the difference of the resummed and fixed order rate 
indicates the size of the higher order corrections, 
we conclude that for inclusive (pseudo-)scalar production 
the $O(\alpha _s^2)$ corrections are likely much smaller than
the uncertainty  from the parton distribution functions 
(cf. Fig.\ref{fig:InitKFac}).  

In Fig.~\ref{fig:QTDistn}, we present the numerical results for the
transverse momentum distributions of the charged top-pions (in TopC model)
and the charged Higgs bosons (in 2HDM) produced at the
upgraded Tevatron and the LHC. The solid curves show the resummation
prediction for the typical values of $m_\phi$. The dashed curves, from 
the $O(\alpha _s)$ prediction, are irregular as $Q_T \to 0$. The large
difference of the transverse momentum distributions
between the results from the resummation and fixed-order analyses 
throughout a wide range of $Q_T$ shows the importance of using
the resummation prediction when extracting the top-pion and Higgs boson 
signals. 
We also note that the average value of $Q_T$ varies slowly as $m_\phi$
increases and it ranges from 
35 to 51\,GeV for $m_\phi$ between 250 and 550 GeV at the 14 TeV LHC, and from 
23 to 45\,GeV for $m_\phi$ between 200 and 300 GeV at the  2 TeV Tevatron.
\pagebreak
\begin{figure}[HT]
\begin{center}
\vspace*{-1cm}
\hspace*{-1.2cm}
\includegraphics[width=18cm,height=17cm]{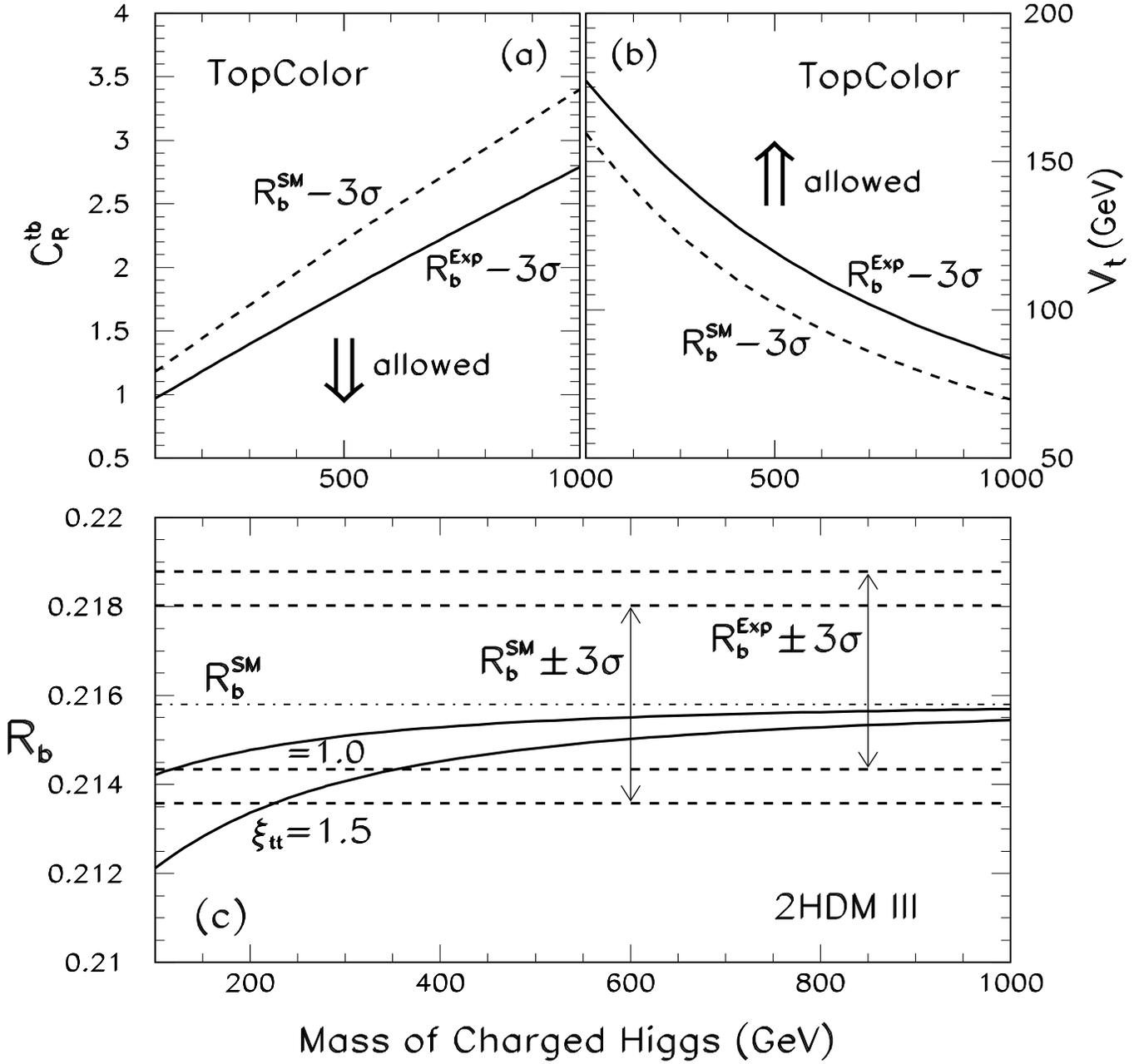}
\end{center}
\vspace*{-0.5cm}
\caption{\small 
Estimated current $3\sigma$-bounds in the TopC model and 2HDM-III:
(a) the $3\sigma$ upper bound on the top-pion Yukawa coupling 
${\cal C}_R^{tb}$; (b) the $3\sigma$ lower bound on the top-pion
decay constant; (Here, in (a) and (b), the solid curves are derived
from the combined LEP/SLD data of 
$R_b^{\rm Exp}=0.21656\pm 0.00074$ while dashed
curves are from the same $3\sigma$ combined experimental error but with
the central $R_b$-value equal to $R_b^{\rm SM}=0.2158$.)
(c) the $R_b$-predictions of 2HDM-III with coupling $\xi_{tt}=1.0$ and
$1.5$ (solid curves) and the $3\sigma$ $R_b$-bounds (dashed lines).
}
\label{fig:Rb}
\end{figure}
\clearpage

\begin{figure}[ht]
\begin{center}
\includegraphics[width=16cm,height=12cm]{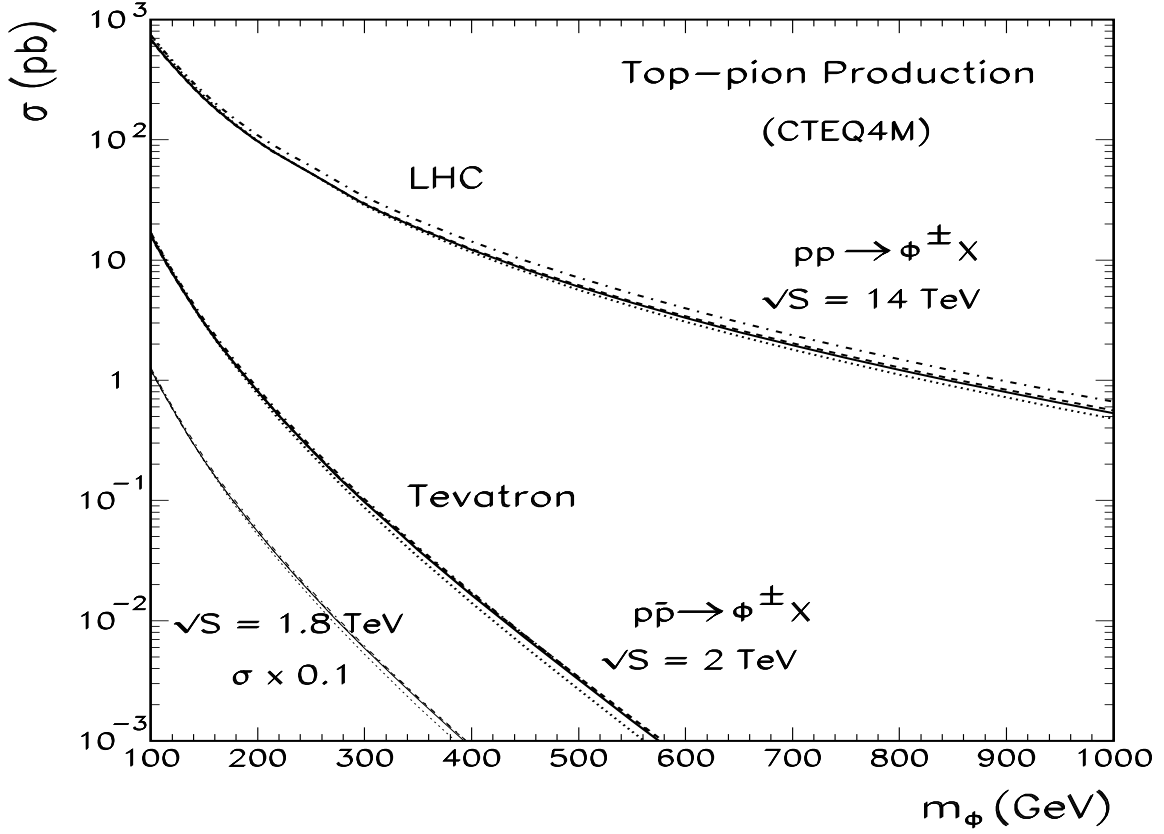}
\end{center}
\vspace{-0.5cm}
\caption{\small 
Top-pion production cross sections at the present Tevatron, upgraded
Tevatron, and the LHC. For each collider we show the NLO cross section with
the resummed running Yukawa coupling (solid), and with one-loop
Yukawa coupling (dashed), as well as the
LO cross section with resummed running Yukawa coupling (dotted) and 
with tree-level (dash-dotted) Yukawa coupling.
}
\label{fig:FigSigma_FixMt.eps}
\end{figure}
\begin{figure}[H]
\begin{center}
\vspace*{-9mm}
\includegraphics[width=16cm,height=12cm]{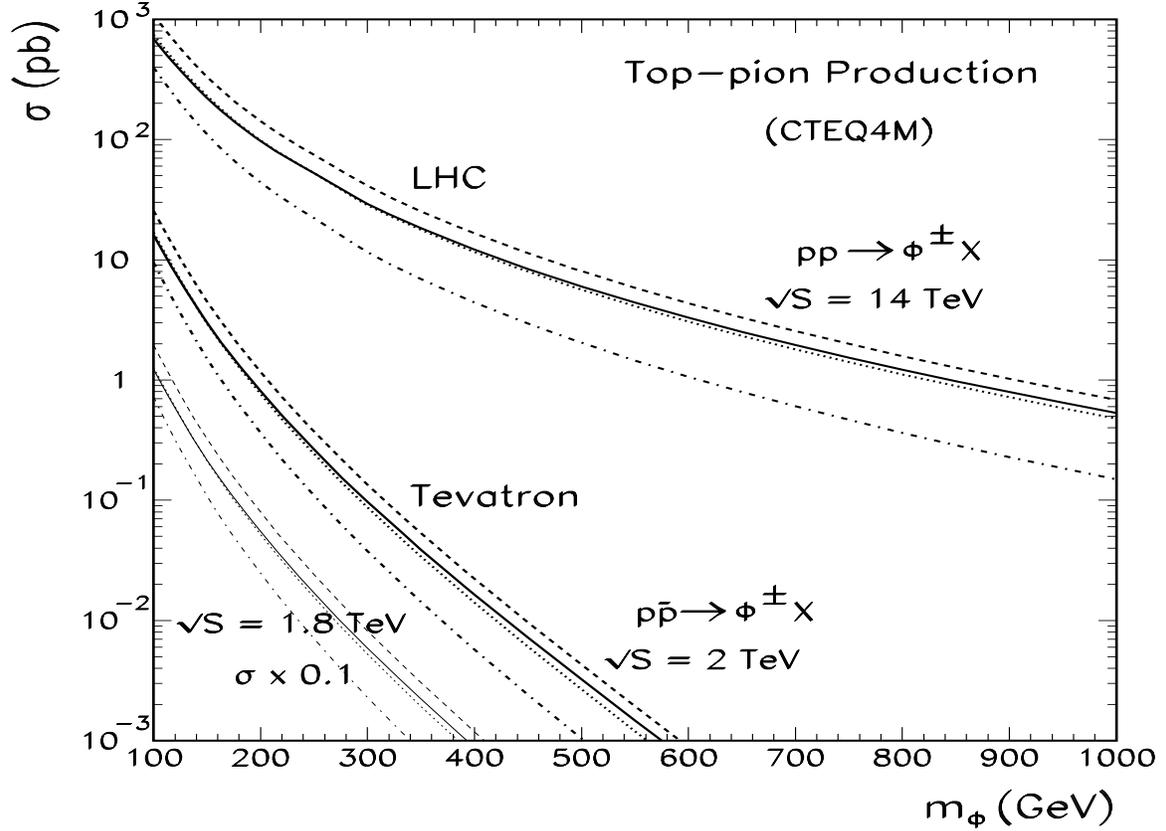}
\vspace*{-7mm}
\end{center}
\caption{\small 
Cross sections for the
charged top-pion production in the TopC model at the present Tevatron,
upgraded Tevatron and the LHC. 
The NLO (solid), the $q\bar{q}'$ (dashed) and $qg$
(dash-dotted) sub-contributions, and the LO (dotted) contributions are shown.
Since the $qg$ cross sections are negative, they are multiplied by
$-1$ in the plot. The cross sections at $\sqrt{S} = 1.8$~TeV are 
multiplied by 0.1 to avoid overlap with the $\sqrt{S} = 2$ TeV curves.   
}
\label{fig:Sigma1}
\end{figure}
\begin{figure}[H]
\begin{center}
\hspace*{-0.8cm}
\vspace*{-7mm}
\includegraphics[width=17.5cm,height=16cm]{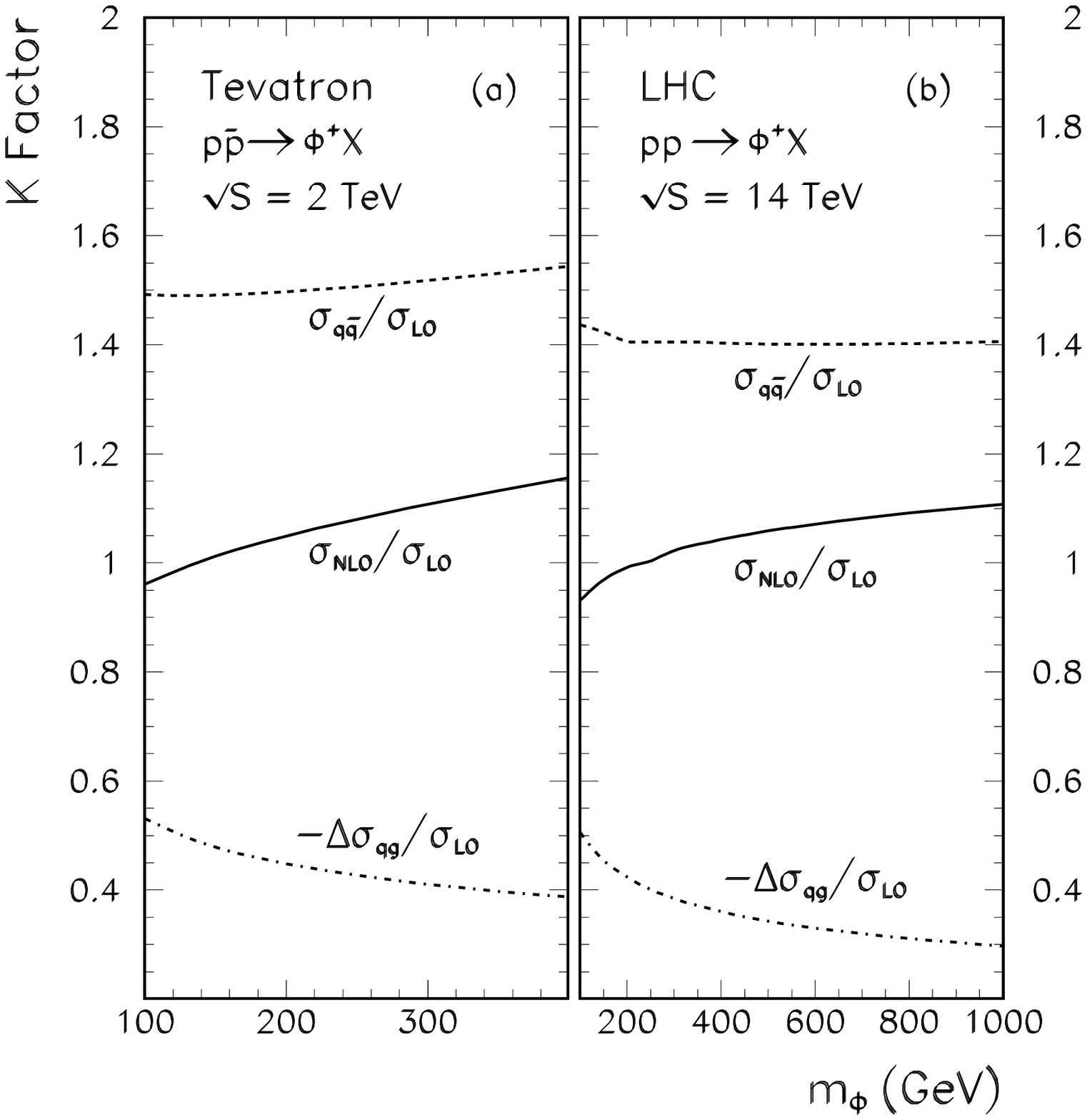}
\end{center}
\caption{\small 
The $K$-factors for the $\phi^+$ production in the TopC model are shown for
the NLO ($K=\sigma_{\rm NLO}/\sigma_{\rm LO}$, solid lines),
$q\bar{q}'$ ($K=\sigma_{\qqbar '}/\sigma_{\rm LO}=
(\sigma_{\rm LO}+\Delta\sigma_{\qqbar '})/\sigma_{\rm LO}$, 
dashed lines), and $qg$ ($K=-\Delta\sigma_{qg}/\sigma_{\rm LO}$, 
dash-dotted lines) contributions, 
at the upgraded Tevatron (a) and the LHC (b).
}
\label{fig:InitKFac}
\end{figure}
\begin{figure}[ht]
\begin{center}
\hspace*{-0.6cm}
\vspace*{-1.2cm}
\includegraphics[width=17cm,height=16cm]{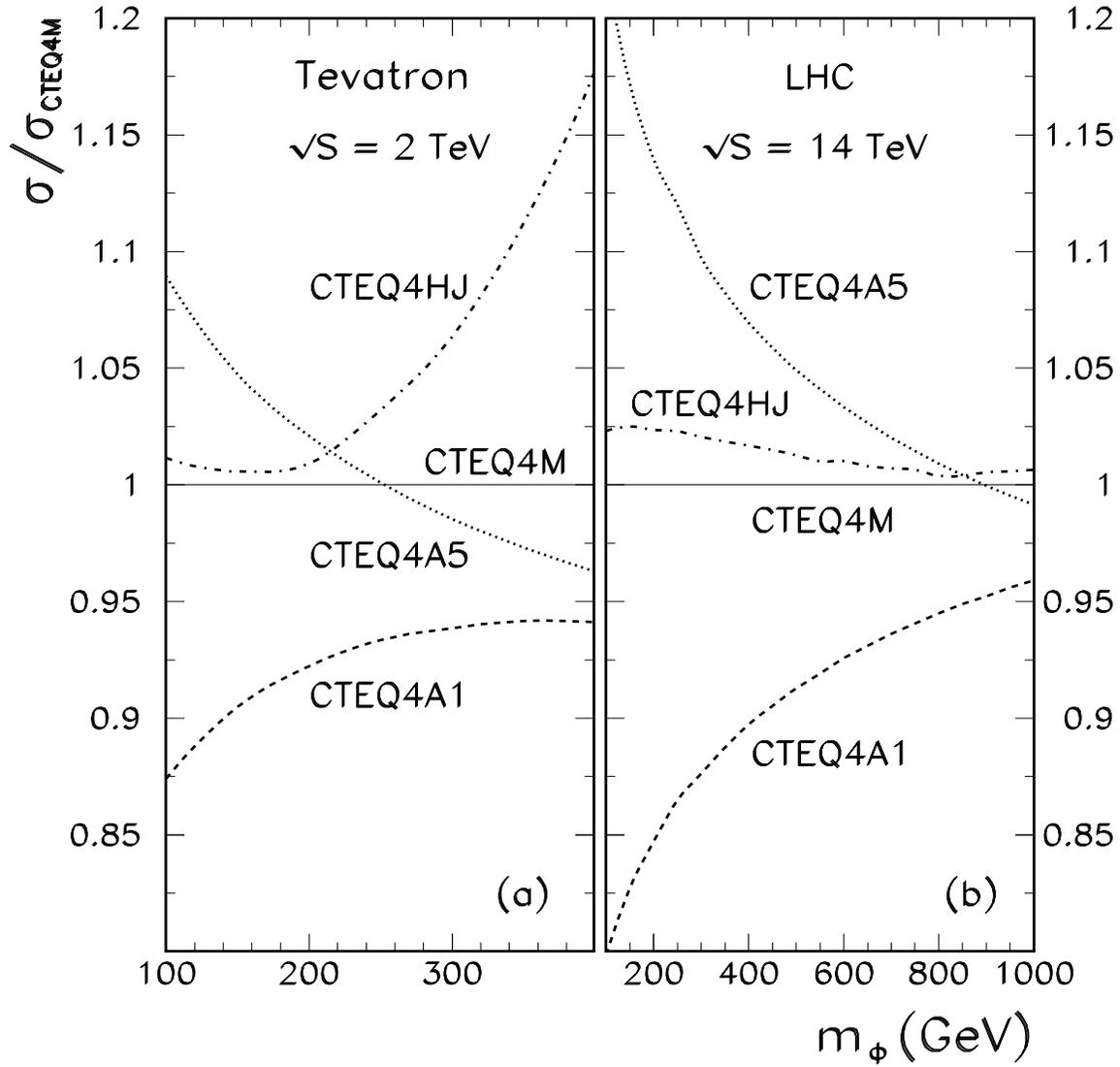}
\vspace*{8mm}
\end{center}
\caption{\small
The ratios of NLO cross sections computed by four different sets of 
CTEQ4 PDFs relative to that by the CTEQ4M
for charged top-pion production 
at the upgraded Tevatron (a) and the LHC (b). 
}
\label{fig:PDF}
\end{figure}
\begin{figure}[H]
\vspace*{-5mm}
\begin{center}
\includegraphics[width=16cm,height=13.6cm]{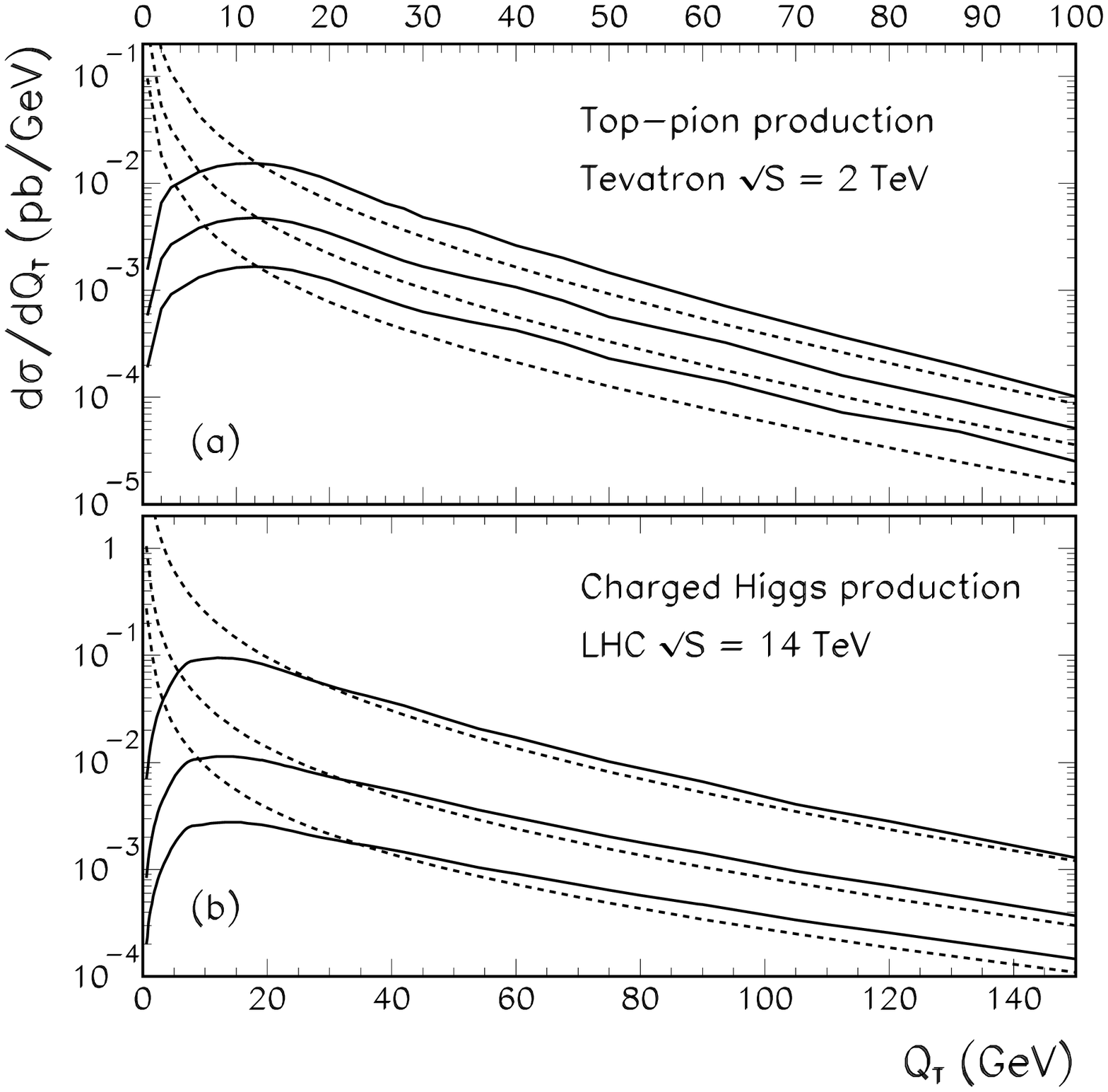}
\end{center}
\vspace*{-4.5mm}
\caption{\small 
Transverse momentum distributions of charged top-pions produced in hadronic
collisions. The resummed (solid) and $O(\alpha _s)$ (dashed)
curves are calculated for $m_\phi =$ 200, 250, and 300 GeV at the
upgraded Tevatron (a), 
and for $m_\phi =$ 250, 400, and 550 GeV at the LHC (b). 
}
\label{fig:QTDistn}
\end{figure}

\vspace{0.5cm}
\noindent
{\normalsize\bf  3.~Hadronic Decays of Charged Scalars 
                    to $O(\alpha_s)$}
\vspace{0.3cm}

In the TopC models, the current Tevatron data from the top quark decay into
charged top-pion ($\pi_t^\pm$) and $b$-quark already requires the
mass of $\pi_t^\pm$ to be above $\sim $150~GeV \cite{TopC,bound}. 
In the current analysis, we shall consider $m_{\pi_t} > m_t+m_b$, 
so that its dominant decay channels are $\pi_t^\pm\to tb,cb$\,.

The decay width of $\pi_t^\pm$ ($=\phi^\pm$), 
including the $O(\alphas )$ QCD corrections, 
is given by \cite{Htb-NLO,CSL}:

\be
\ba{l}
\dis\Gamma_{\rm NLO}(Q)~=~\dis \Gamma_{\rm LO}(Q)
\[1+\f{\alphas C_F}{2\pi}{\cal R}\],~~~~~\\[3mm]
\Gamma_{\rm LO}(Q) ~= 
\dis\f{3}{16\pi}Q\( |{\cal C}_L|^2+|{\cal C}_R|^2\)
(1-r)^2,\\[2.8mm]
{\cal R} ~=~\dis\f{9}{2}(1-r)^2+(1-r)\(3-7r+2r^2\)\ln\f{r}{1-r}
 +\[\dis 3\ln\f{Q^2}{m_t^2}+4-\Omega\]\\[2.3mm]
 ~~~~~~~~\dis -2(1-r)^2\[\f{\ln (1-r)}{1-r}-2{\rm Li}_2\(\f{r}{1-r}\)
-\ln (1-r)\ln\f{r}{1-r} \],  
\ea
\label{eq:widNLO}
\ee
in which $Q=\sqrt{Q^2}$ is the invariant mass of $\phi^\pm$.
The small bottom and charm masses are ignored so that
~$r\equiv \( m_t/m_{\phi}\)^2$ for $tb$ final state and $~r=0~$
for $cb$ final state.  Thus, for $~\phi^\pm \to cb$,~ the quantity
${\cal R}$ reduces to $\,{\cal R} = {17}/{2}-\Omega$\,.
In Fig.~\ref{fig:GammaBr}, we present the results for
total decay widths of $\phi^+$ and branching ratios of $\phi^+ \to t\bar{b}$
in the TopC model and the 2HDM. For the 2HDM, we also show the branching 
ratios of the $W^+h^0$ channel, which is complementary to the $t\bbar$
channel.  The NLO (solid) and LO (dashed) curves differ only by a small
amount. In the same figure,
the $K$-factor, defined as the ratio of the NLO to the LO partial decay 
widths, is plotted for the  $\phi^+\to t\bar{b}$ (solid) 
and $\to c\bar{b}$ (dashed) channels. Here,
the sample results for the 2HDM  are derived for the parameter 
choice: $\alpha =0$ and $(M_h,\,M_A)=(100,\,1200)$\,GeV.

With the decay width given above, we can study the invariant
mass distribution of $\,tb\,$ for the $s$-channel $\phi^+$-production:
\be
\dis\f{d\sigma}{dQ^2}
\[h_1h_2\to\hspace*{-0.7mm}(\phi^+X)\hspace*{-0.7mm}\to t\bbar X\]
=\dis\sigma\[h_1h_2\hspace*{-1.5mm}\to\hspace*{-0.7mm}\phi^+(Q)X\] 
\f{\(Q^2\Gamma_\phi/m_\phi\){\rm Br}\[\phi^+\hspace*{-1.5mm}\to t\bbar\]}
{\pi\[\(Q^2\hspace*{-0.7mm}-m_\phi^2\)^2+\(Q^2\Gamma_\phi
/m_\phi\)^2\]},
\label{eq:M_tb}
\ee
where $\Gamma_\phi$ and ${\rm Br}\[\phi^+\hspace*{-1.5mm}\to t\bbar\]$
are the total decay width of $\phi^+$ and the branching ratio of
$\phi^+\to t\bbar$, respectively, which are calculated up to the NLO.
We note that the one-loop box diagrams 
with a virtual gluon connecting the initial state quark and
final state quark (from the hadronic decay of $\phi$)
have vanishing contribution at $O(\alphas)$
because the scalar $\phi$ is color-neutral.
In Fig.~\ref{fig:DsDQ}a and Fig.~\ref{fig:DsDQ-LHC}a,
we plot the invariant mass distribution for 
$t$-$\bar b$ and $\bar t$-$b$ pairs from $\phi^\pm$ (top-pion signal) and
$W^{\pm\ast}$ (background) decays in the TopC model. In these plots,
we have included the NLO contributions, as a function of $Q$, 
to the $W^{\pm \ast}$ background rate at the Tevatron and the LHC. 
The overall $K$-factor (after averaging over the invariant mass $Q$) 
including both the initial and final state radiations is about 1.4~(1.34) 
for the Tevatron (LHC)\,\cite{Tim-CP}.
The total rate of $W^{\pm \ast}$ up to the NLO is about 0.70 [0.86]~pb and 
11.0~pb at the 1.8 [2]~TeV Tevatron and the 14~TeV LHC, respectively.
\begin{figure}[ht]
\begin{center}
\vspace*{-1cm}
\includegraphics[width=16cm,height=17cm]{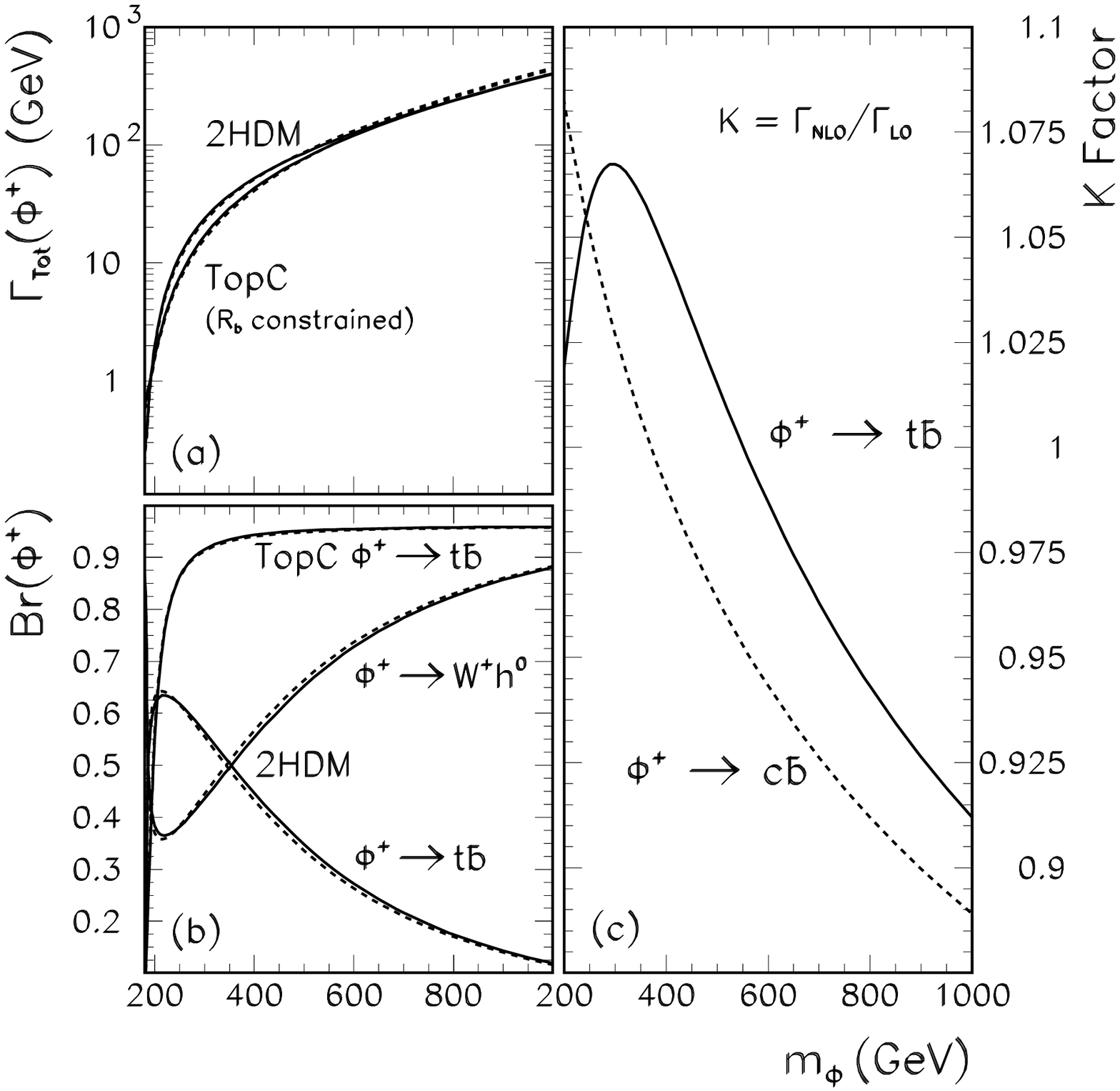}
\end{center}
\vspace*{-0.5cm}
\caption{\small 
Total decay widths of $\phi^+$ and BRs of $\phi^+ \to t\bar{b}$ in the
TopC model and 2HDM. (For the 2HDM, the BR of the $W^+h^0$ channel is
also shown, which is complementary to the $t\bbar$ channel.) In Fig. (a)
and (b), the NLO (solid) and LO (dashed) curves differ only by a small
amount. In Fig. (c), the $K$-factor, which is defined as the ratio of the
NLO to the LO partial decay widths, is shown for the $\phi^+\to
t\bar{b}$ (solid) and $\to c\bar{b}$ (dashed) channels. The sample
results for the 2HDM in this figure are derived for the parameter choice
$(\xi_{tt}^U,\,\xi_{tc}^U)=(1.5,\,1.5)$, 
$\alpha =0$, and $(m_h,\,m_A)=(120,\,1200)$\,GeV.
}
\label{fig:GammaBr}
\end{figure}

\begin{figure}[ht]
\vspace*{-1.7cm}
\begin{center}
\includegraphics[width=16cm,height=21cm]{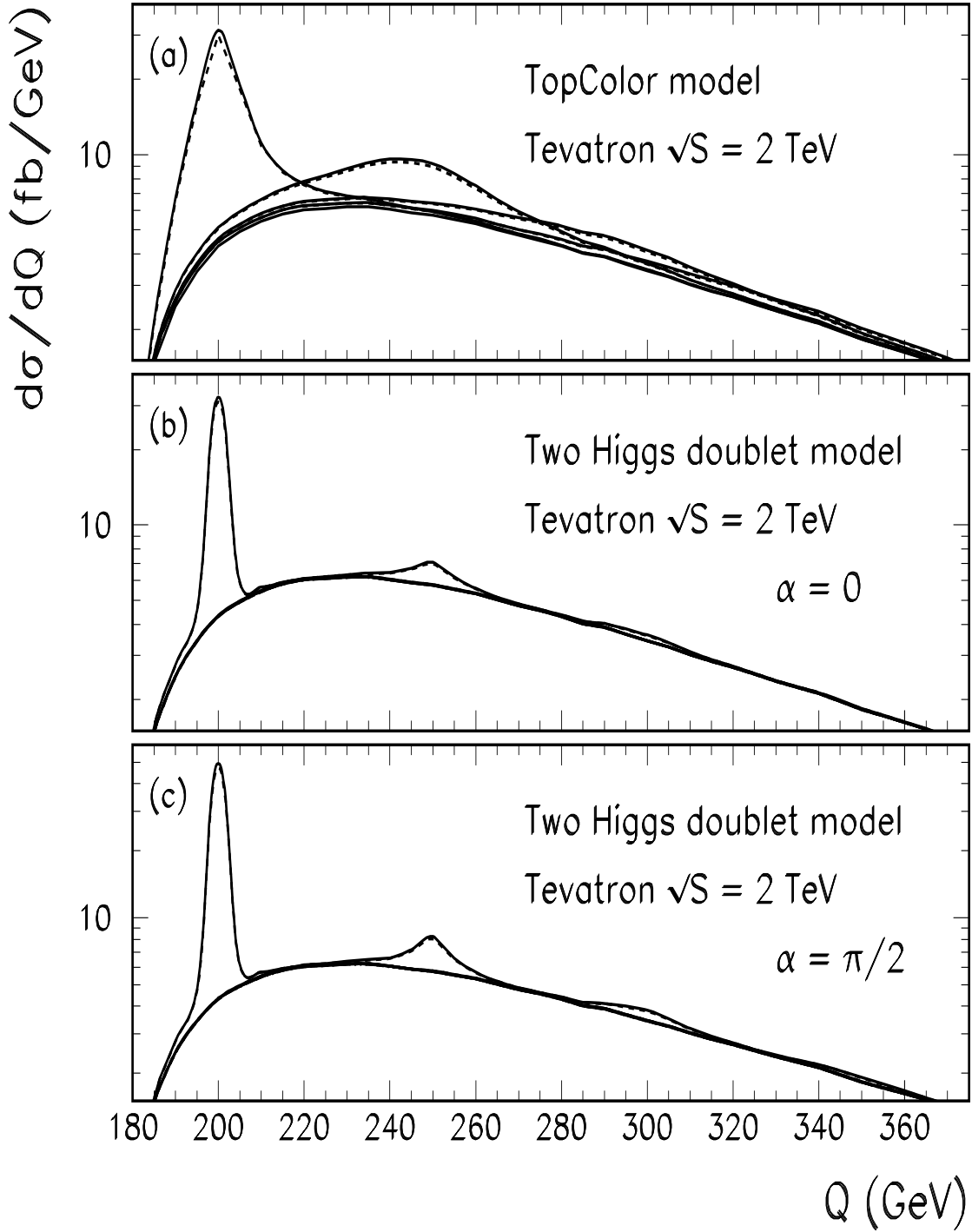}
\end{center}
\vspace*{-0.7cm}
\caption{\small
Invariant mass distribution of $t$-$\bar{b}$ and  $\bar{t}$-$b$
pairs from  $\phi^\pm$ (signal) and $W^{\pm *}$ (background) 
decays at the Tevatron Run-II for the
TopC model (a), and 2HDM with Higgs mixing angles $\alpha =0$ (b),
and $\alpha =\pi /2$ (c). 
We show the signal for $m_{\phi}=200,~250$, 300 and 350\,GeV. The solid
curves show the results from the NLO calculation,
and the dashed ones from the LO analysis. 
}
\label{fig:DsDQ}
\end{figure}
\clearpage
\begin{figure}[ht]
\vspace*{-1.7cm}
\begin{center}
\includegraphics[width=16cm,height=21cm]{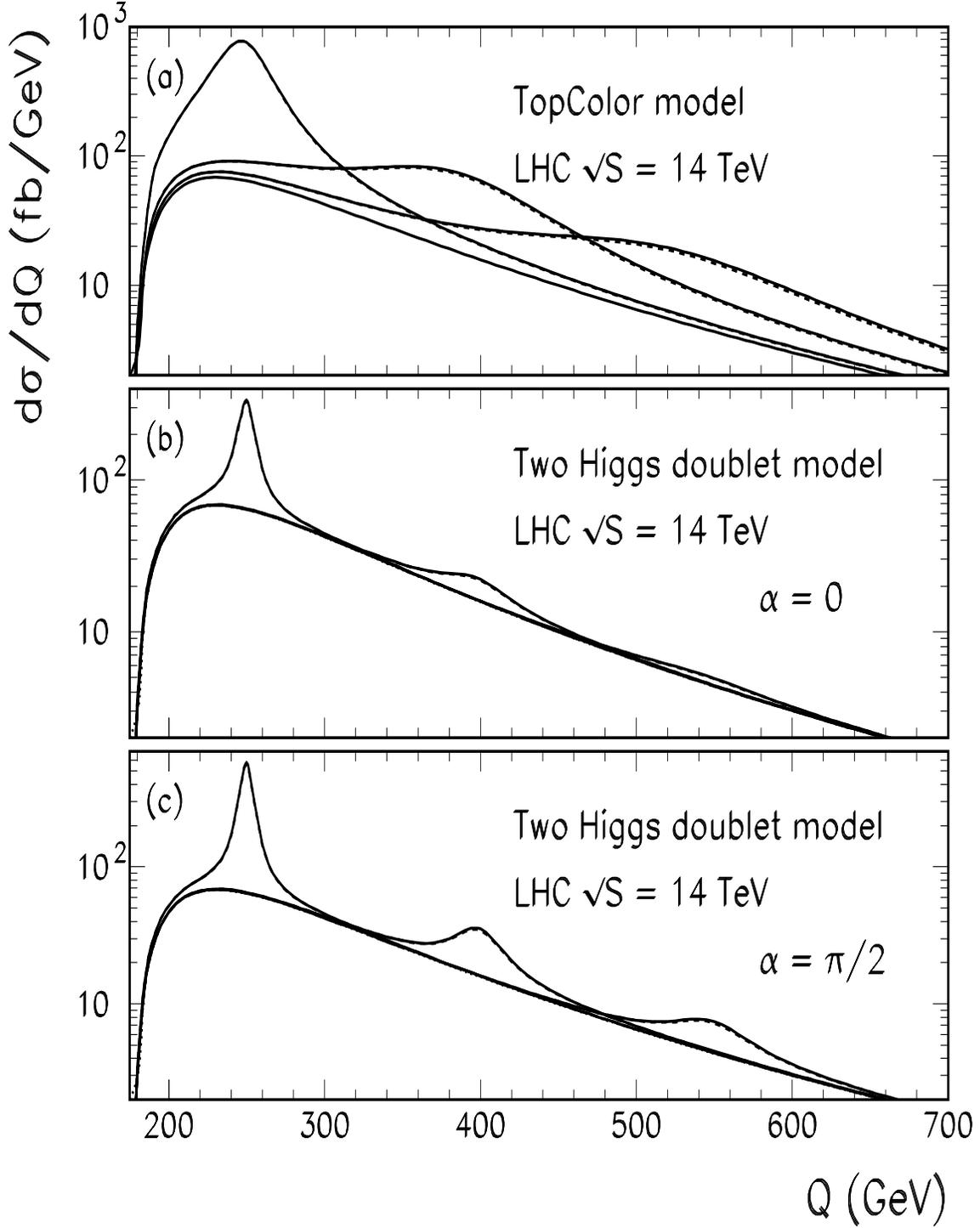}
\end{center}
\vspace*{-0.7cm}
\caption{\small
Invariant mass distribution of $t$-$\bar{b}$ and $\bar{t}$-$b$ pairs
from $\phi^\pm$ (signal) and $W^{\pm *}$ (background) decays at the LHC
for the TopC model (a), and for the 2HDM with the Higgs mixing angles
$\alpha =0$ in (b), and $\alpha =\pi/2$ in (c). Here the charged
pseudo-scalar or scalar mass are chosen as the typical values of
$m_{\phi}=250,~400$ and 550\,GeV. The solid curves show the results by
the NLO calculation, while the dashed ones come from the LO analysis. 
}
\label{fig:DsDQ-LHC}
\end{figure}
\clearpage

\begin{figure}[ht]
\vspace*{-0.5cm}
\begin{center}
\includegraphics[width=16cm,height=14cm]{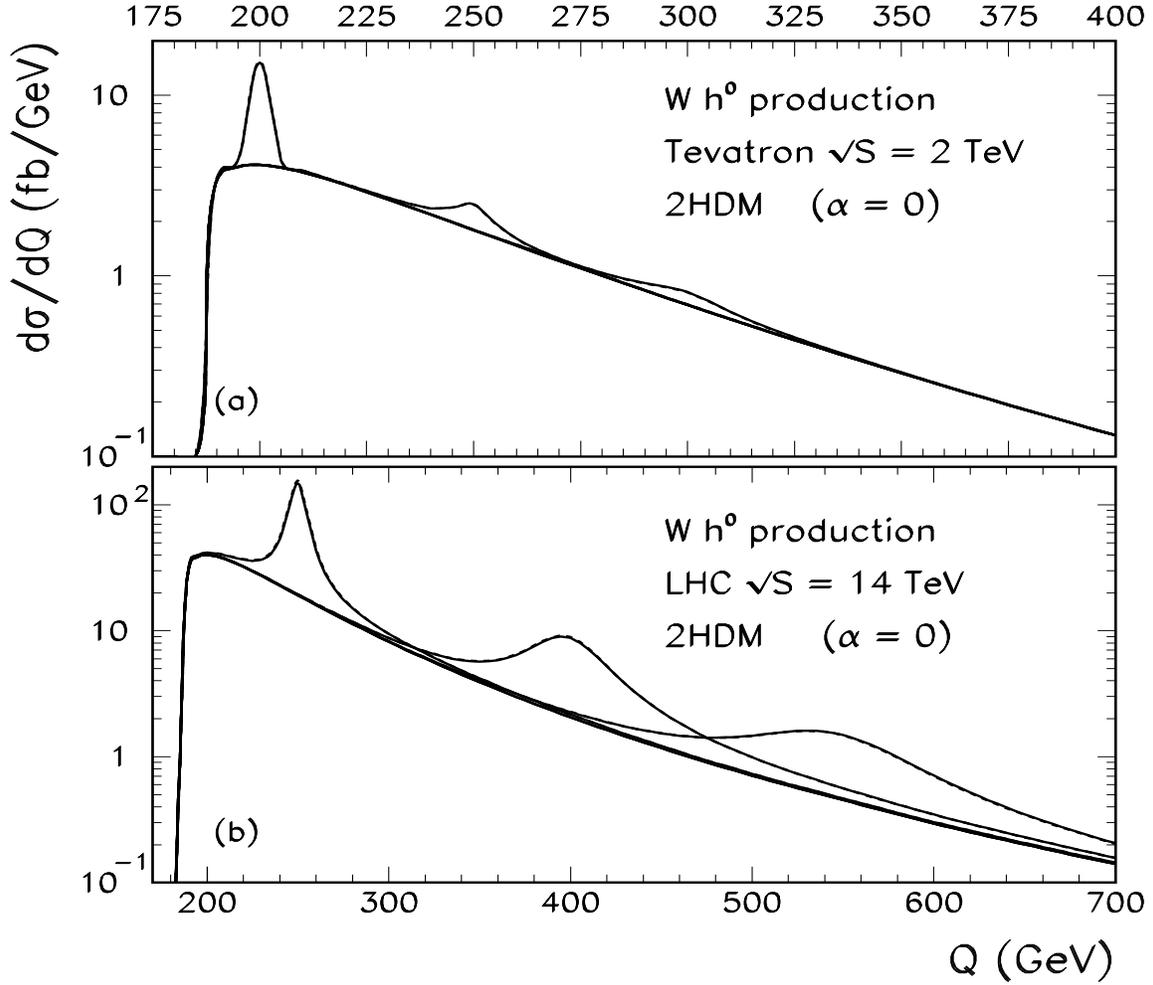}
\end{center}
\vspace*{-0.7cm}
\caption{\small
Invariant mass distributions of $W^+$-$h^0$ and $W^-$-$h^0$ pairs from
$\phi^\pm$ ($s$-channel resonance) and 
$W^{\pm *}$ ($s$-channel non-resonance) decays at the Tevatron
Run-II, and at the LHC, for the 2HDM with Higgs mixing angles
$\alpha=0$. We show the signal for $m_{\phi}=200,~250$ and 300\,GeV at
the Tevatron (a), and for $m_{\phi}=250,~400$ and 550\,GeV at the LHC (b). The
solid curves show the results of the NLO calculation, and the dashed
ones of the LO analysis. 
}
\label{fig:DsDQ-Wh}
\end{figure}

Before concluding this section, we discuss how to generalize the above
results to the generic 2HDM (called type-III \cite{2HDM3}), 
in which the two Higgs doublets $\Phi_1$
and $\Phi_2$ couple to both up- and down-type quarks and the {\it ad hoc}
discrete symmetry \cite{GW} is not imposed.
The flavor-mixing Yukawa couplings in this model can be conveniently
formulated under a proper basis of Higgs doublets so that
{\small $~<\hspace*{-1.2mm}\Phi_1\hspace*{-1.2mm}>=(0,v/\sq2 )^T$} and
{\small $<\hspace*{-1.2mm}\Phi_2\hspace*{-1.2mm}>=(0,0)^T$.}
Thus, the diagonalization of the fermion mass matrix also 
diagonalizes the Yukawa couplings of $\Phi_1$,  
and all the flavor-mixing effects are generated by Yukawa couplings 
($\widehat{Y}^U_{ij}$ and $\widehat{Y}^D_{ij}$) 
of $\Phi_2$ which exhibit a natural hierarchy under the 
ansatz \cite{ansatz,2HDM3}
\be
\dis \widehat{Y}_{ij}^{U,D}=\xi_{ij}^{U,D}
{\sqrt{m_im_j}}/
{<\hspace*{-1.2mm}\Phi_1\hspace*{-1.2mm}>}
\label{eq:ansatz}
\ee
with {\small $~\xi_{ij}^{U,D}\sim O(1)$}.~
This ansatz highly suppresses the flavor-mixings among light quarks
and identifies the largest mixing coupling as the one from the
$t$-$c$ or $c$-$t$ transition. 
A recent renormalization group analysis~\cite{RG-2HDM3}
shows that such a suppression persists at the high energy scales.
The relevant Yukawa interactions involving the charged Higgs bosons
$H^\pm$ are \cite{HY}:
\be 
\ba{ll}
{\cal L}_Y^{CC} & =~ H^+\[\overline{t_R}~(\widehat{Y}_U^\dag V)_{tb}~b_L
        -\overline{t_L}~(V\widehat{Y}_D)_{tb}~b_R  
        ~+\overline{c_R}~(\widehat{Y}_U^\dag V)_{cb}~b_L
        -\overline{c_L}~(V\widehat{Y}_D)_{cb}~b_R~\]+{\rm h.c.}
\\[2mm]
 & \simeq~ H^+\[\overline{t_R}~\widehat{Y}^{U\ast}_{tt}~b_L
               +\overline{c_R}~\widehat{Y}^{U\ast}_{tc}~b_L\]
               +{\rm h.c.}+({\rm small~terms})\,,
\ea
\label{eq:Hcb}
\ee
where $ ~\widehat{Y}^U_{tt}=\xi^U_{tt}\times (\sqrt{2}m_t/v)\simeq \xi^U_{tt}$,
and  $~\widehat{Y}^U_{tc}=\xi^U_{tc}\times (\sqrt{2m_tm_c}/v)
                  \simeq  \xi^U_{tc}\times 9\%~$,
in which $~\xi^U_{tc}\sim O(1)$ is allowed by 
the current low energy data \cite{2HDM3,Laura}.
As a result, the Yukawa counter term in Fig.~1d involves both $\delta m_t$ and
$\delta m_c$. Consequently, we need to replace the NLO quantity $\Omega$ 
in the finite part of the Yukawa counter term 
[cf. the definition below (\ref{eq:NLO})] by 
\be
\dis\Omega ({\rm 2HDM}) = 3\ln \left[m_\phi^2/(m_tm_c)\right]+4~, 
\label{eq:omega-2HDM}
\ee
for the type-III 2HDM.
In the relevant $\phi^\pm$-$c$-$b$ coupling of this 2HDM, we note that, 
similar to the case of the TopC model,
only the right-handed charm is involved \cite{HY}, i.e.,
\be
{\cal C}_L^{tb}={\cal C}_L^{cb}= 0,~~~~
{\cal C}_R^{tb}=\xi_{tt}^U(\sqrt{2}m_t/v),~~~
{\cal C}_R^{cb}\simeq \xi^U_{tc}\times 9\% .
\label{eq:choice0}
\ee
where the parameters ($\xi^U_{tt},\,\xi_{tc}^U$) 
are expected to be naturally
around $O(1)$. We have examined the possible constraint of $\xi^U_{tt}$
from the current $R_b$ data and found that the values of
$\xi^U_{tt}\!\sim\! 1.0\!-\!1.5$ are allowed for 
$m_{H^\pm}\gtrsim 200$\,GeV (cf. Fig.\,\ref{fig:Rb}c).\footnote{
Our calculation of $R_b$ in the 2HDM-III is consistent with those
in Ref.\,\cite{Rb2HDM} and Ref.\,\cite{Laura} after using the same inputs. 
Note that a larger value of $\xi_{tt}^U$ than ours was chosen for the
solid curve in Fig.~3 of Ref.\,\cite{Laura}.
We thank L. Reina for clarifying the inputs of Ref.\,\cite{Laura} and for
useful discussions.}\,  
The production cross section of $H^\pm$
in this 2HDM can be obtained by rescaling
the result of the TopC model according to the ratio of the coupling-square
$\[{\cal C}^{tc}_R({\rm 2HDM})/{\cal C}_R^{tc}({\rm TopC})\]^2
\!\sim\! \[0.09\xi_{tc}^U/0.2{\cal C}_R^{tb}({\rm TopC})\]^2$ (which is
about $1/7$ for $\xi_{tc}^U=1.5$ and
the charged scalar mass around $400$\,GeV). 

Finally, we note that there are three neutral Higgs bosons 
in the 2HDM, the CP-even scalars ($h^0,~H^0$) 
and the CP-odd pseudo-scalar $A^0$. The mass diagonalization
for $h^0$ and $H^0$ induces the Higgs mixing angle $\alpha$. 
The low energy constraints on this model require \cite{2HDM3,Laura}
$~m_h,\,m_H \leq m_{H^\pm} \leq m_A~$ or 
$~m_A \leq m_{H^\pm} \leq m_h,\,m_H $.~   
For the case of
$m_{H^\pm} > m_{h^0} + M_W$, the $H^\pm\to W^\pm h^0$ decay channel is also
open. Taking, for example,  
$\alpha =0$ and $(m_h,\,m_A)=(120,1200)$\,GeV, we find from
Fig.~\ref{fig:GammaBr}b that the $tb$ and $Wh^0$ decay 
modes are complementary at low and
high mass regions of the charged Higgs boson $H^\pm$. 
In Figs.~\ref{fig:DsDQ}b-c and \ref{fig:DsDQ-LHC}b-c,
we plot the invariant mass distributions of $t\bbar$ 
and $\tbar b$ pairs from $H^\pm$ (signal) and $W^{\pm\ast}$ (background)
decays in the 2HDM  at the 2\,TeV Tevatron and the 14\,TeV LHC,
with the typical choice of the parameters: 
$~\(\xi^U_{tt},\,\xi^U_{tc}\)=(1.5,\,1.5)$ in  Eq.~(\ref{eq:choice0}), 
$\(m_h,\,m_A\)=\(120,\,1200\)$\,GeV, and $\alpha=0$ or $\pi/2$.
[A larger value of $\xi^U_{tt}$ will
simutaneously increase (reduce) the BR of $tb$ ($Wh^0$) mode.]
We see that, due to a smaller $c$-$b$-$H^\pm$ coupling  
[cf. (\ref{eq:choice0})], it is hard to detect such a charged Higgs boson
with mass $m_{H^\pm}>250$\,GeV at the Tevatron Run-II. We then examine
the potential of the LHC for the high mass range of $H^\pm$. Similar
plots are shown in Figs.~\ref{fig:DsDQ}b-c for 
$\alpha =0$ and $\alpha =\pi/2$, respectively. 
When $\cos\alpha$ is large (e.g., $\alpha =0$),
the branching ratio of the $tb$-channel decreases as $m_{H^\pm}$ 
increases (cf. Fig.~\ref{fig:GammaBr}b),  so that the LHC does not 
significantly improve the probe of the large $m_{H^\pm}$, 
range via the single-top mode (cf. Fig.~\ref{fig:DsDQ-LHC}b).
In this case,  the $W^\pm h^0$ channel, however,
becomes important for large $m_{H^\pm}$, as shown in Fig.~\ref{fig:DsDQ-Wh}
(cf. Fig.~\ref{fig:GammaBr}b, for its decay branching ratios) 
since the $H^\pm$-$W^\mp$-$h^0$
coupling is proportional to $\cos \alpha$ \cite{2HDM3}.
On the other hand, for the parameter space with
small $\cos\alpha$ (e.g., $\alpha =\pi /2$), the $W^\pm h^0$ channel 
is suppressed so that the single-top mode is important even for large 
mass region of $H^\pm$.\footnote{Note that the $H^\pm$-$W^\mp$-$H^0$
coupling is proportional to $\sin\alpha$ and is thus enhanced for
small $\cos\alpha$. In this case, the $WH^0$ mode may be important
provided that $H^0$ is relatively light. We will not further elaborate
this point here since it largely depends on the mass of $H^0$.}\,
This is illustrated in Fig.~\ref{fig:DsDQ-LHC}c at the LHC for
$\alpha =\pi/2$. In order to probe the whole parameter space and
larger $m_{H^\pm}$, it is important to study both $tb$ 
and $Wh^0$ (or $WH^0$) channels.

\vspace*{0.5cm}
\noindent
{\normalsize\bf 4.~Generalization to Neutral Scalar Production
via $b\bbar$ Fusion} 
\vspace*{0.15cm}

The QCD corrections are universal so that the generalization to 
the production of neutral scalar or pseudo-scalar $\phi^0$ via
the $b\bbar$ fusion is straightforward,
i.e., we only need to replace (\ref{eq:omega-2HDM}) by 
\be
\dis\Omega (\phi^0 b\bbar ) = 3\ln \left[m_\phi^2/m_b^2\right]+4\,, 
\label{eq:omega-Hbb}
\ee
in which $m_\phi$ is the mass of $\phi^0$.
The finite piece of the Yukawa renormalization
[cf. the quantity $\Omega$ in (\ref{eq:delta_mt})]
is scheme-dependent.
We can always define the $\phi^0$-$b$-$\bbar$ Yukawa coupling as 
$\sqrt{2}m_b/v$ times an enhancement factor $K$ so that the Yukawa
counter term is generated by $\delta m_b/m_b$.\footnote{
This specific definition works even if the Yukawa coupling is not related to 
any quark mass. For instance,
the bottom Yukawa couplings of the $b$-Higgs and
$b$-pion in the TopC model \cite{TopC,hbb} are independent of quark masses
because the $b$-Higgs does not develop VEV.} 
After resumming the leading logarithmic terms,
$\[\alpha_s \ln (m_\phi^2/m_b^2)\]^n$, via the renormalization
group technique, the net effect of the Yukawa renormalization is to
change the Yukawa coupling or the related quark-mass into the corresponding
$\overline{\rm MS}$ running coupling or mass, as discussed in the previous
section.

The $b\bbar$ decay branching ratios of the neutral Higgs bosons in 
the MSSM with large $\tanb$
are almost equal to one \cite{Hdecay}. The same is
true for the $b$-Higgs or $b$-pion in the TopC model \cite{TopC}.
It has been shown that at the Tevatron,
the $b\bbar$ dijet final states can be properly identified \cite{CDF_bb}.
The same technique developed for studying the resonance of the
coloron or techni-$\rho$ in the $b\bbar$ decay mode \cite{CDF_bb} can
also be applied to the search of the neutral Higgs bosons 
with large bottom Yukawa coupling.
When the neutral scalar or pseudo-scalar $\phi^0$ 
is relatively heavy, e.g., in the range of $O(250-1000)$~GeV, 
the QCD dijet backgrounds can be effectively removed by requiring the two
$b$-jets to be tagged with large transverse momenta ($P_T$) because 
the $P_T$ of each $b$-jet from the $\phi^0$ decay 
is typically at the order of $m_\phi/2$.
Hence, this process can provide complementary information to that 
obtained from studying 
the $\phi^0 b\bbar$ associate production \cite{Jack,hbb,hbx}.

We first consider the production of the
neutral Higgs boson $\phi^0$, which can be either $A^0$, $h^0$, or $H^0$,
in the MSSM with large $\tanb$, where the corresponding Yukawa couplings
to $b\bbar$ and $\tau^+\tau^-$ 
are enhanced relative to that of the SM since 
$y_{D}/y_{D}^{\rm SM}$ is equal to $\tanb$, $-\sin\alpha/\cos\beta$, or
$\cos\alpha/\cos\beta$, respectively, at the tree-level.
In the large $\tanb$ region,
the MSSM neutral Higgs bosons dominantly decay into $b\bbar$ and
$\tau^+\tau^-$ final states,
which can be detected at the hadron colliders. In comparison with
the recent studies on the $\phi^0b\bbar$ \cite{hbb} 
and $\phi^0\tau^+\tau^-$ \cite{htautau} 
associate production, we expect the inclusive $\phi^0$
production via the $b\bbar$-fusion would be more useful for $m_\phi$ being
relatively heavy (e.g., $m_\phi \geq 200-300$~GeV) 
because of the much larger phase space as well as
a better suppression of the backgrounds in the high $P_T$ region. 
The total LO and NLO cross sections 
for the inclusive production process $pp,p\bar{p}\to A^0X$ at the
Tevatron and the LHC are shown in Figs.~\ref{fig:Sigma_MSSMbb}a and b, 
in parallel to Figs.~\ref{fig:FigSigma_FixMt.eps} and \ref{fig:Sigma1}
for the case of charged top-pion production.
Here, we have chosen \,$\tanb =40$\, for illustration. The cross sections
at other values of $\tanb$ can be obtained by multiplying the scaling factor
$\, \(\tanb /40\)^2 \,$.
From Fig.~\ref{fig:Sigma_MSSMbb}a, we see a significant improvement from
the pure LO results (dash-dotted curves) by 
resumming over the large logarithms of $m_\phi^2/m_b^2$ into the running
Yukawa coupling.  The good agreement
between the LO results with running Yukawa coupling and the NLO results
is due to a non-trivial, and process-dependent,
cancellation between the individual $O(\alpha_s)$ contributions 
of the $b\bbar$ and $bg$ sub-processes.
In contrast to the production of the charged top-pion or Higgs boson
via the initial state $c\bbar$ or $\cbar b$ partons, the neutral Higgs 
boson production involves the $b\bar{b}$ parton densities. 
The $K$-factors for the ratios of the NLO versus LO cross sections of
$p\bar{p}/pp\to A^0X$ are presented in Fig.~\ref{fig:K-A0} for the MSSM 
with $\tanb =40$. The main difference is due to 
the fact that the individual contribution
by the $O(\alpha_s)$ $bg$-fusion becomes more negative as compared to
the case of the charged top-pion production shown in Fig.~\ref{fig:InitKFac}. 
This makes the overall $K$-factor of the NLO versus LO cross sections
range from about $-(16$$\sim$17)\% to +5\% at the Tevatron and the LHC.
In parallel to Table~1 and Fig.~\ref{fig:PDF}, 
we have examined the uncertainties of the CTEQ4 PDFs
for the $A^0$-production at the Tevatron and the LHC, and the results are
summarized in Table~\ref{tb:PDFS2} and Fig.~\ref{fig:PDF-bb}).  
We also note that, similar to the charged Higgs boson production, 
the resummed total rate for the neutral Higgs boson production is
not very different from its NLO rate.

The transverse momentum ($Q_T$) distributions of $A^0$,
produced at the upgraded Tevatron and at the LHC, are shown in
Fig.~\ref{fig:QT-MSSM} for various $A^0$ masses ($m_A$) with $\tan \beta = 40$.
The solid curves are the result of the multiple soft-gluon resummation,
and the dashed ones are from the $O(\alpha_s)$ calculation.
The shape of these transverse momentum distributions is similar to that
of the charged top-pion (cf. Fig.~\ref{fig:QTDistn}). 
The fixed order distributions are singular as
$Q_T \to 0$, while the resummed ones have a maximum at some finite $Q_T$
and vanish at $Q_T = 0$. When $Q_T$ becomes large, of the order of $m_A$, 
the resummed curves merge into the fixed order ones. The average resummed
$Q_T$ varies between 25 and 30 (40 and 60) GeV in the 
mass range of $m_A$ from 200 to 300 (250 to 550)\,GeV
at the Tevatron (LHC).

We also note that for large $\tanb$,
the SUSY correction to the running $\phi^0$-$b$-$\bbar$ Yukawa coupling
is significant \cite{large-tanb} and can be
included in a way similar to our recent analysis of the $\phi^0b\bbar$
associate production~\cite{hbb}.  
To illustrate the SUSY correction to the $b$-Yukawa coupling, we choose all 
MSSM soft-breaking parameters as $500$~GeV, and the Higgs mixing parameter
$\mu = \pm 500$~GeV. Depending on the sign of $\mu$, the SUSY correction 
to the $\phi^0$-$b$-${\bar b}$ coupling can either take the same sign as 
the QCD correction or have an opposite sign \cite{hbb}.
In Fig.~\ref{fig:Sigma_MSSMbb}c, the solid curves represent the NLO
cross sections with QCD correction alone, while the results including the 
SUSY corrections to the running bottom Yukawa coupling are shown for
$\mu =+500$\,GeV (upper dashed curves) and $\mu =-500$\,GeV (lower dashed
curves). As shown, these partial SUSY corrections can change the cross
sections by about a factor of 2.
The above results are for the inclusive production of the CP-odd Higgs 
boson $A^0$ in the MSSM. Similar results can be easily obtained for the
other neutral Higgs bosons ($h^0$ and $H^0$) by properly rescaling
the coupling strength. We also note that in the large $\tanb$ region,
there is always a good mass-degeneracy between either $h^0$ and $A^0$ (in 
the low mass region with $m_A \lae 120$\,GeV) or $H^0$ and $A^0$ (in
the high mass region with $m_A \gae 120$\,GeV), as shown in Figs.~10 and 11
of Ref.~\cite{hbb}.
\begin{table}[ht]
\caption{\small 
Cross sections in fb for neutral Higgs boson production in the MSSM 
with $\tanb =40$, at the upgraded Tevatron and the LHC, are shown
for four different CTEQ4 PDFs. 
They are separately given for the LO and NLO processes, 
and for the $b\bar{b}\to A^0X$ and $bg\to A^0X$ sub-processes. 
For the upgraded Tevatron the 
top number is for $m_{A} = 200$ GeV, 
the middle is for $m_{A} = 300$ GeV, and
the lowest is for $m_{A} = 400$ GeV.
For the LHC the 
top number is for $m_{A} = 400$ GeV, 
the middle is for $m_{A} = 700$ GeV, and
the lowest is for $m_{A} = 1$ TeV.
}
\vspace*{4mm}
\begin{center}
\begin{tabular}{c||r r r r|r r r r}
\hline\hline
&&&&&&&&\\[-0.2cm]
Collider & 
\multicolumn{4}{c|}{Upgraded Tevatron (2\,TeV)} 
& \multicolumn{4}{c}{LHC (14\,TeV)} \\
[0.15cm] \cline{1-9} &&&&&&&&\\[-0.2cm]
Process $\backslash$ PDF 
&   4A1 &    4M &   4A5 &   4HJ &   4A1 &    4M &   4A5 &   4HJ \\
[0.15cm]\hline\hline &&&&&&&&\\[-0.2cm]
&  2020 &  1900 &  1660 &  1920 & 18100 & 19800 & 16600 & 17900 \\
&&&&&&&&\\[-0.3cm] LO                  
&   166 &   153 &   129 &   163 &  1520 &  1440 &  1280 &  1440 \\
&&&&&&&&\\[-0.3cm]
&  19.9 &  18.2 &  15.0 &  21.7 &   258 &   238 &   206 &   238 \\
[0.2cm]\hline &&&&&&&&\\[-0.2cm]
&  1810 &  1780 &  1620 &  1800 & 17100 & 17400 & 16700 & 17500 \\
&&&&&&&&\\[-0.3cm] NLO                 
&   160 &   154 &   134 &   164 &  1520 &  1470 &  1350 &  1470 \\
&&&&&&&&\\[-0.3cm]
&  20.3 &  19.3 &  16.4 &  22.9 &   265 &   250 &   222 &   251 \\
[0.2cm]\hline &&&&&&&&\\[-0.2cm]
&  3040 &  2900 &  2590 &  2930 & 25400 & 25400 & 24100 & 25600 \\
&&&&&&&&\\[-0.3cm] $q\bar{q}\to\phi^0X$
&   253 &   237 &   203 &   251 &  2140 &  2050 &  1850 &  2050 \\
&&&&&&&&\\[-0.3cm]
&  31.0 &  28.8 &  24.0 &  33.8 &   364 &   339 &   298 &   340 \\
[0.2cm]\hline &&&&&&&&\\[-0.2cm]
&$-$1230&$-$1120&$-$970&$-$1130&$-$8320&$-$8010&$-$7370&$-$8050\\
&&&&&&&&\\[-0.3cm] $qg\to\phi^0X$      
&$-$92.9&$-$83.1&$-$69.0&$-$87.5& $-$623& $-$575& $-$505& $-$574\\
&&&&&&&&\\[-0.3cm]
&$-$10.6&$-$9.42&$-$7.59&$-$10.9& $-$98.8&$-$88.8&$-$75.8&$-$88.7\\
[0.2cm]\hline\hline
\end{tabular}
\end{center}
\label{tb:PDFS2}
\end{table}
\begin{figure}[ht]
\vspace*{-2cm}
\begin{center}
\includegraphics[width=16cm,height=18cm]{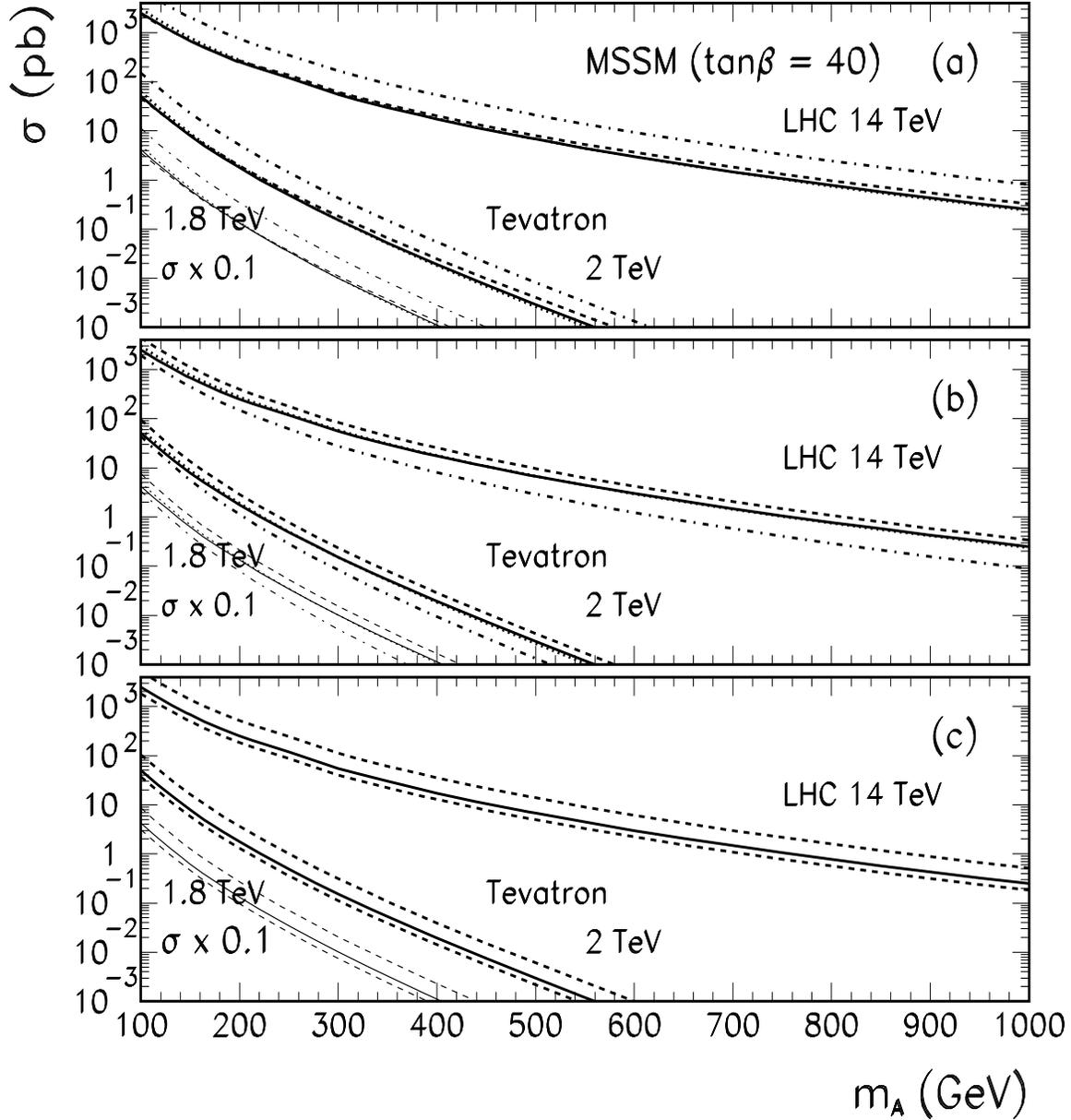}
\end{center}
\vspace*{-7mm}
\caption{\small 
LO and NLO cross sections for the neutral Higgs $A^0$ 
production in the MSSM with $\tanb =40$,
at the Tevatron and the LHC. 
(a) For each collider we show the NLO cross sections with
the resummed running Yukawa coupling (solid) and with one-loop
Yukawa coupling (dashed), as well as the
LO cross sections with resummed running Yukawa coupling (dotted) and 
with tree-level Yukawa coupling (dash-dotted).
(b) The NLO (solid), the $b\bar b$ (dashed) and $bg$
(dash-dotted) sub-contributions, and the LO (dotted) contributions are shown.
Since the $bg$ cross sections are negative, they are multiplied by
$-1$ in the plot. The cross sections at $\sqrt{S} = 1.8$~TeV are 
multiplied by 0.1 to avoid overlap with the $\sqrt{S} = 2$ TeV curves.   
(c) The NLO cross sections
with QCD running Yukawa coupling (solid curves) and those with additional
SUSY correction to the running coupling are shown 
(upper dashed lines for the Higgs-mixing parameter $\mu =+500$\,GeV 
and lower dashed lines for $\mu =-500$\,GeV).
}
\label{fig:Sigma_MSSMbb}
\end{figure}
\clearpage
\begin{figure}[ht]
\begin{center}
\vspace*{-7mm}
\includegraphics[width=16cm,height=12cm]{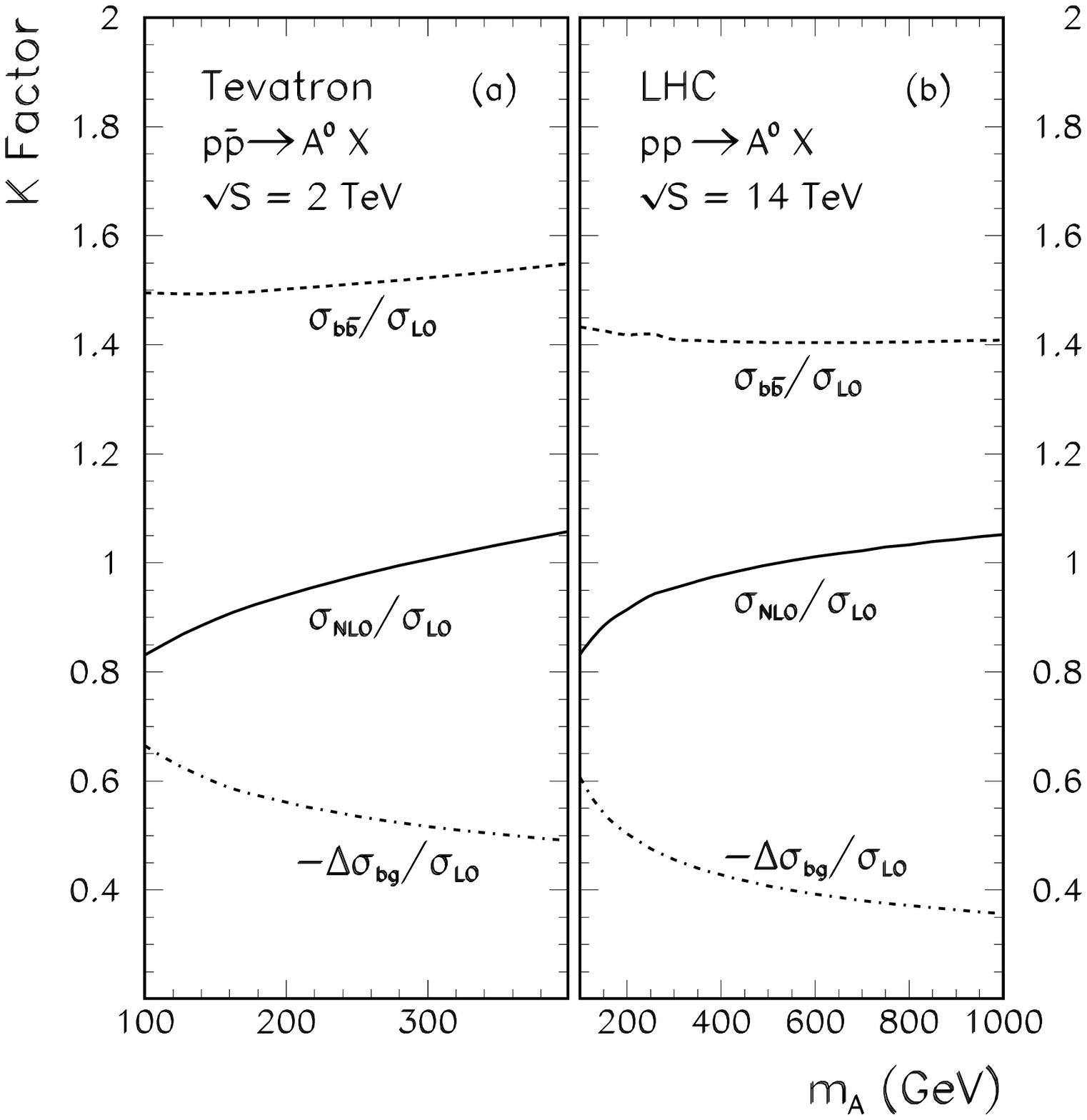}
\vspace*{-7mm}
\end{center}
\caption{\small 
The $K$-factors for the $A^0$ production in the MSSM with 
$\tanb =40$ are shown for
the NLO ($K=\sigma_{\rm NLO}/\sigma_{\rm LO}$, solid lines),
$b\bar{b}$ ($K=\sigma_{\bb}/\sigma_{\rm LO}=
(\sigma_{\rm LO}+\Delta\sigma_{\bb})/\sigma_{\rm LO}$, 
dashed lines), and $bg$ ($K=-\Delta\sigma_{bg}/\sigma_{\rm LO}$, 
dash-dotted lines) contributions, 
at the upgraded Tevatron (a) and the LHC (b).
}
\label{fig:K-A0}
\end{figure}
\begin{figure}[ht]
\begin{center}
\vspace*{-1.2cm}
\includegraphics[width=16cm,height=14cm]{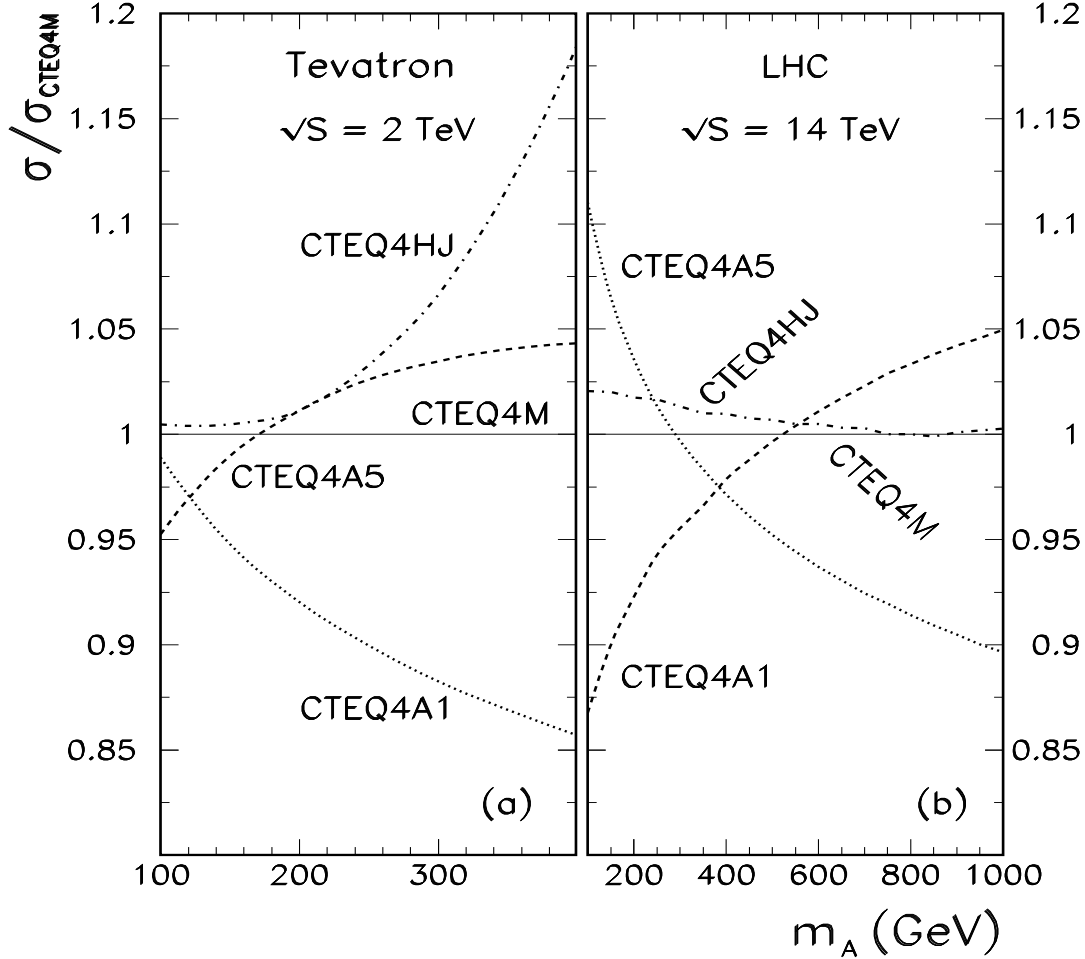}
\vspace*{-3mm}
\end{center}
\caption{\small
The ratios of NLO cross sections computed by four different sets of 
CTEQ4 PDFs relative to that by the CTEQ4M
for neutral $A^0$-production in the MSSM with $\tanb =40$, 
at the upgraded Tevatron (a) and the LHC (b). 
}
\label{fig:PDF-bb}
\end{figure}
\begin{figure}[ht]
\begin{center}
\vspace*{-1.2cm}
\includegraphics[width=16cm,height=14cm]{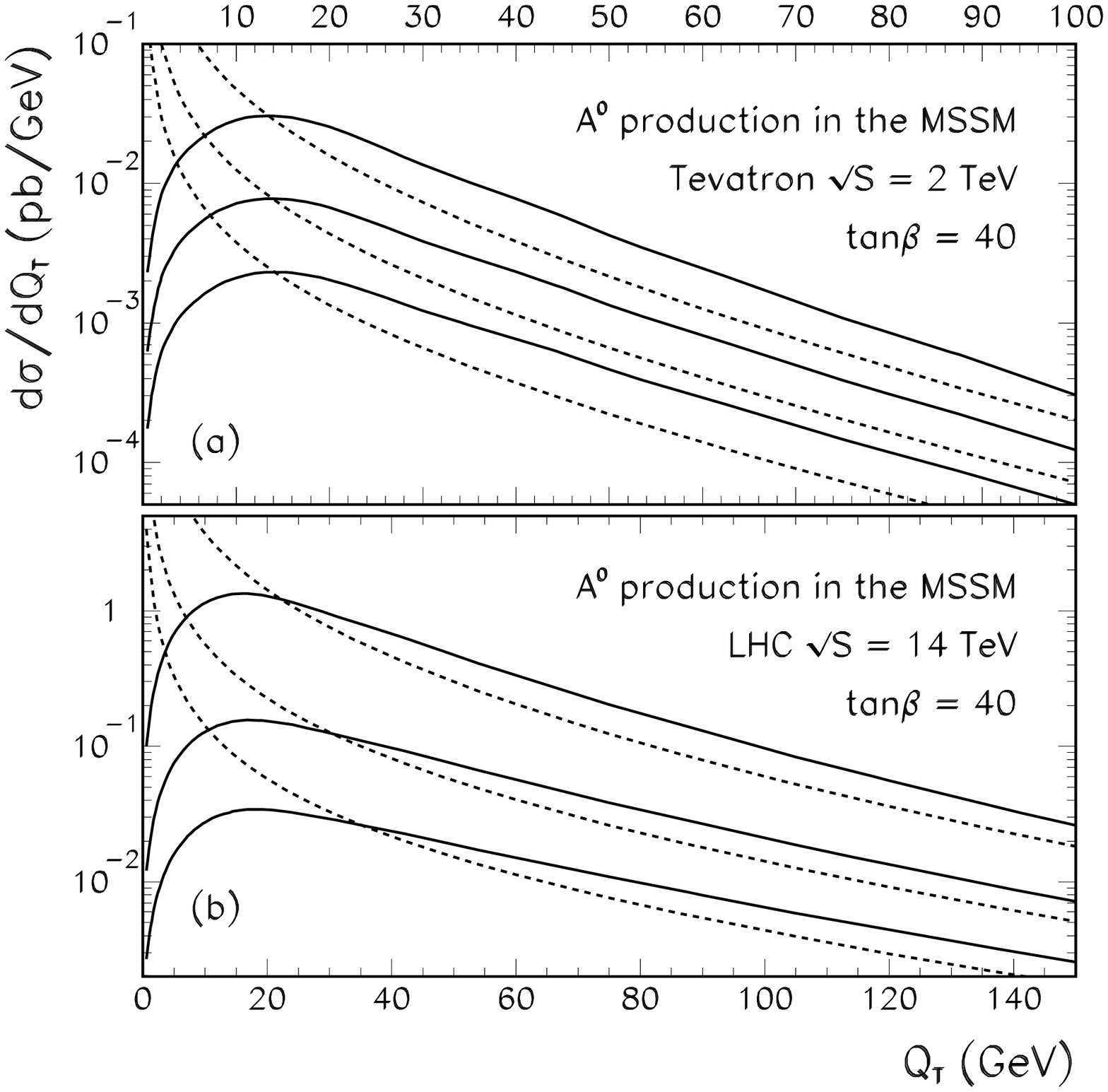}
\vspace*{-3mm}
\end{center}
\caption{\small
Transverse momentum distributions of pseudo-scalar $A^0$ produced
via hadronic collisions, calculated in the MSSM with $\tanb =40$. 
The resummed (solid) and $O(\alpha _s)$ (dashed) curves 
are shown for $m_A =$ 200, 250, and
300\,GeV at the upgraded Tevatron (a), and 
for $m_A =$ 250, 400, and 550\,GeV at the LHC (b).
}
\label{fig:QT-MSSM}
\end{figure}

We then consider the large bottom Yukawa coupling of
the neutral $b$-Higgs ($h_b^0$) and $b$-pion
($\pi_b^0$) in the TopC model \cite{TopC,TopC2,hbb}.
The new strong $U(1)$ force in this model is attractive in the 
$<\hspace*{-0.15cm}\tbar t\hspace*{-0.15cm}>$ channel but repulsive 
in the $<\hspace*{-0.15cm}\bbar b\hspace*{-0.15cm}>$ channel. Thus, 
the top but not the bottom acquires dynamical mass from the vacuum. 
This makes the  $t$-Yukawa coupling ($y_t$) super-critical
while the $b$-Yukawa coupling ($y_b$) sub-critical, at the TopC breaking
scale $\Lambda$, i.e., 
~$y_b(\Lambda ){~\lae ~} y_{\rm crit}
\hspace*{-1mm}=\hspace*{-1mm}\sqrt{{8\pi^2}/{3}} 
{~\lae ~} y_t(\Lambda ) $~,   which requires
$y_b$ being close to $y_t$ and thus naturally large. Our recent
renormalization group analysis \cite{hbb} shows that the relation
$y_b(\mu ){\sim } y_t(\mu )$ holds well at any scale $\mu$ below $\Lambda$.
For the current numerical analysis, we shall choose a typical value of 
~$y_b(m_t)\simeq y_t(m_t)\approx 3$,~ i.e., 
~$|{\cal C}_L^{bb}|=|{\cal C}_R^{bb}|\simeq 3/\sqrt{2}.~$
In Fig.~\ref{fig:Sigma-bHiggs}, we plot the production cross sections of 
$h^0_b$ or $\pi_b^0$ at the Tevatron and the LHC.  This is similar to the
charged top-pion production in Fig.~\ref{fig:Sigma1}, except the non-trivial
differences in the Yukawa couplings (due to the different tree-level values 
and the running behaviors) and the charm versus bottom parton luminosities.
\begin{figure}[ht]
\begin{center}
\vspace*{-9mm}
\includegraphics[width=16cm,height=12cm]{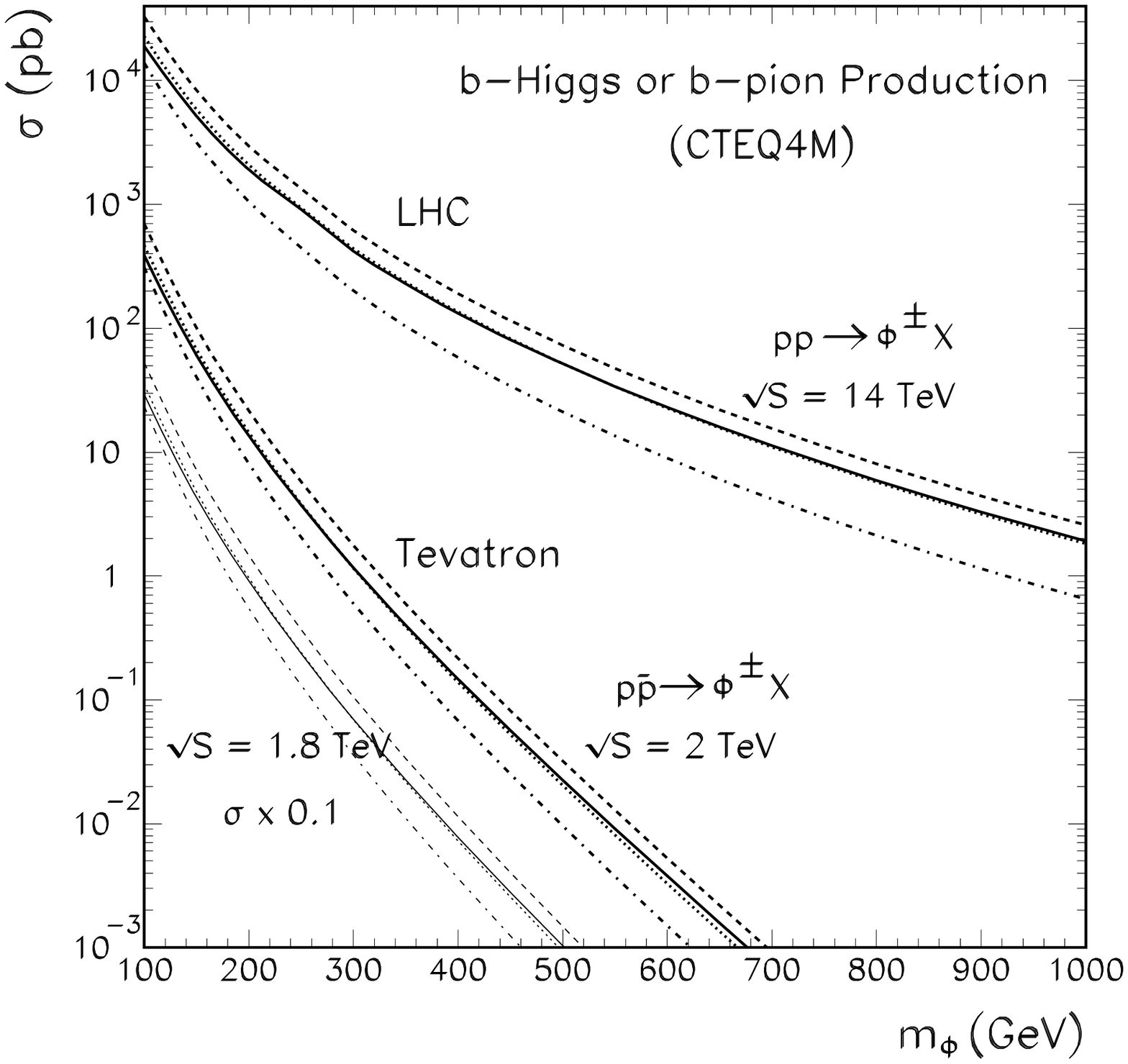}
\vspace*{-7mm}
\end{center}
\caption{\small
Cross sections for the neutral $b$-pion $\pi_b^0$ or $b$-Higgs $h_b^0$
production via the $b\bbar$-fusion in the TopC model at the Tevatron and
the LHC. The NLO (solid), the $q\bar{q}'$ (dashed) and $qg$
(dash-dotted) sub-contributions, and the LO (dotted) contributions with
resummed running Yukawa coupling are shown. Since the $qg$ cross
sections are negative, they are multiplied by $-1$ in the plot. The cross
sections at $\sqrt{S} = 1.8$~TeV are multiplied by 0.1 to avoid overlap
with the $\sqrt{S} = 2$ TeV curves. 
}
\label{fig:Sigma-bHiggs}
\end{figure}

\vspace*{0.5cm}
\noindent
{\normalsize\bf 5.~Conclusions}
\vspace*{0.3cm}

In summary, we have presented the complete 
$O(\alpha_s)$ QCD corrections
to the charged scalar or pseudo-scalar production via the
partonic heavy quark fusion process at hadron colliders. 
We found that 
the overall NLO corrections to the $p\bar{p}/pp \to \phi^\pm$
processes are positive for $m_\phi$ above $\sim$150\,(200)~GeV
and lie below $\sim$15\,(10)\% for the
Tevatron (LHC) in the relevant range of $m_\phi$
(cf.  Fig.~\ref{fig:InitKFac}). 
The inclusion of the NLO contributions thus
justifies and improves our recent LO analysis \cite{HY}.
The uncertainties of the NLO rates
due to the different PDFs are systematically examined and are found to be
around $20\%$ (cf. Table~1 and Fig.~\ref{fig:PDF}).  
The QCD resummation to include the effects of multiple soft-gluon 
radiation is also performed,
which provides a better prediction of the transverse momentum ($Q_T$)
distribution of the scalar $\phi^{0,\pm}$, 
and is important for extracting the experimental signals 
(cf. Fig.~\ref{fig:QTDistn}).
We find that the resummed total rate differs from the $O(\alpha_s)$
rate only by a few percents which indicate that the size of the 
higher order corrections are likely much smaller than the uncertainty 
from the parton distribution functions (cf. Figs.\,6 and 14).
We confirm that the 2\,TeV Tevatron (with a $2\!-\!10$~fb$^{-1}$ 
integrated luminosity) is able to explore the natural 
mass range of the top-pions up to about 300--350\,GeV
in the TopC model \cite{TopC,TopC2} for the typical $t_R$-$c_R$
mixing of $K^{tc}_{UR}\!\sim\! 0.2\!-0.33$ [cf. eq.\,(\ref{eq:KURtc})].   
Measuring the top polarization in the single-top event will further
improve the signal identification.  On the other hand,
due to a possibly smaller $\phi^\pm$-$b$-$c$ coupling in the
2HDM, we show that to probe the charged Higgs boson with mass
above 200\,GeV in this model may require a high luminosity Tevatron 
(with a $10-30$~fb$^{-1}$ integrated luminosity). 
The LHC will further probe the charged Higgs 
boson of the 2HDM up to about $O(1)\,$TeV via the single-top and
$W^\pm h^0$ (or $W^\pm H^0$) production.  
The complementary roles of the $tb$ and $W^\pm h^0$ 
channels in the different regions of the Higgs mass and the Higgs
mixing angle $\alpha$ are demonstrated.   We have also analyzed a
direct extension of our NLO results to the neutral 
(pseudo-)scalar production via the $b\bbar$-fusion for the 
neutral Higgs bosons $(A^0,h^0,H^0)$ in the MSSM with large $\tanb$,
and for the neutral $b$-pion ($\pi_b^0$) or $b$-Higgs ($h_b^0$)
in the TopC model with $U(1)$-tilted large bottom Yukawa coupling.
In comparison with the $\phi^0 b\bbar$ associate production \cite{hbb},
this inclusive $\phi^0$-production mechanism provides a complementary 
probe for a neutral Higgs boson (with relatively large mass),
whose decay products,
e.g., in the $b\bbar$ or $\tau\tau$ channel, typically have
high transverse momenta ($\sim m_\phi /2$) and
can be effectively detected \cite{CDF_bb}.
This is particularly helpful for the discovery reach of the Tevatron. 
Further detailed  Monte Carlo analyses at the detector level 
should be carried out to finally conclude the sensitivity of 
the Tevatron Run-II and the LHC via this process. 
 
At the final stage of writing up this manuscript, 
we became aware of a new preprint \cite{note-added}
which studied the QCD corrections for the neutral
Higgs production $b\bbar \to H^0$ within the SM,  
and partially overlaps with our 
Sec.~4 as the pure NLO QCD correction is concerned. 
The overlapped part is in general agreement with ours except  
that we determine the counter term of the 
Yukawa coupling (expressed in terms of the relevant quark
mass) by the on-shell scheme (cf. Refs.~\cite{Htb-NLO,CSL}) 
while Ref. \cite{note-added} used $\overline{\rm MS}$ scheme. 
After resumming the leading logarithms into
the running mass or Yukawa coupling, the two results coincide.
Note that the apparent large $O(\alphas )$ correction derived in 
Ref.~\cite{note-added} is due to the fact that it only includes
the contribution from the $b\bar{b}$ sub-process, which is 
part of our complete $O(\alphas )$ contribution.
The inclusion of the NLO contribution from the $bg$ sub-process,
which turns out to be negative and partially cancels the $b {\bar b}$ 
contribution, yields a typical size of $O(\alphas )$ correction 
to the production rate of a neutral Higgs boson produced via 
heavy quark fusion. The $bg$ sub-process is identified as 
$O(1/\ln[m_H/m_b])$ instead of $O(\alpha_s)$ correction in 
Ref.~\cite{note-added}.

\vspace*{6mm}
\noindent
{\normalsize\bf Acknowledgments}\\[3mm]     
We thank C.T.~Hill for discussing the top-pion signature at the 
FermiLab Tevatron, H.E.~Haber for discussing the 
charged Higgs production, W.K.~Tung for discussing the heavy quark
parton distribution functions,
and J.~Huston for discussing the CDF 
Run-Ib analysis on the new particle searches via
the $\bb$ dijet mode \cite{CDF_bb}.
We also thank the authors of Ref.~\cite{note-added} 
for confirming our comparison with their results, 
and explaining how they included the $bg$ sub-process.
This work is supported by the U.S.~NSF under grant PHY-9802564.

\newpage
\noindent
{\normalsize\bf Appendix}
\vspace*{4mm}

In this appendix, we present the individual NLO parton cross sections
computed at $D=4-2\epsilon$ dimensions.
We note that, unlike the usual Drell-Yan type processes, the one-loop 
virtual contributions (cf. Fig.~1b-d) 
are not ultraviolet (UV) finite unless the new
counter term from Yukawa coupling (related to the quark-mass renormalization,
cf. Fig.~1e) is included.

\vspace*{1.5mm}
\noindent
{\bf A. Partonic processes $c\bbar\to\phi^+X$}

The spin- and color-averaged amplitude-square for the 
$c\bbar\to\phi^+g$ process is
\be
\overline{|{\cal M}|^2}=\dis
\f{2\pi C_F}{3}\alpha_s\left(|\CL|^2+|\CR|^2\right)\mu^{2\epsilon}
\[(1-\epsilon)\(\f{\widehat{t}}{\widehat{u}}+
\f{\widehat{u}}{\widehat{t}}+2\)+
2\f{\widehat{s}\,m_\phi^2}{\widehat{t}\,\widehat{u}} \].
\label{eq:M2qq}
\ee

The individual contributions 
(from the virtual loop and real gluon emission)
to the NLO partonic cross section are:
\be
\ba{ll}
\Delta\widehat{\sigma}^{\rm virtual}_{\rm loop} \hspace*{-2mm}
&=~\dis\widehat{\sigma}_0\f{\alpha_sC_F}{2\pi}
 \(\f{4\pi\mu^2}{Q^2}\)^{\epsilon}
 \f{\Gamma (1-\epsilon )}{\Gamma (1-2\epsilon )}
 \[ -\f{2}{\epsilon^2}
    +\f{2\pi^2}{3}-2\]\delta (1-\tauhat ),
\\[5mm]
\Delta\widehat{\sigma}^{\rm virtual}_{\rm count} \hspace*{-2mm}
&=~\dis\widehat{\sigma}_0\f{\alpha_sC_F}{2\pi}
 \(4\pi\)^{\epsilon}
 \f{\Gamma (1-\epsilon )}{\Gamma (1-2\epsilon )}
 \[ -\f{3}{\epsilon}-\Omega\]\delta (1-\tauhat ),
\\[5mm]
\Delta\widehat{\sigma}^{\rm real}_{c\bbar} \hspace*{-2mm}
&=~\dis\widehat{\sigma}_0\f{\alpha_sC_F}{2\pi}
\(\f{4\pi\mu^2}{Q^2}\)^{\epsilon}
 \f{\Gamma (1-\epsilon )}{\Gamma (1-2\epsilon )}
\left[ \f{2}{\epsilon^2}\delta (1-\tauhat )
+\f{3}{\epsilon}\delta (1-\tauhat )
-\f{2}{\epsilon}P^{(1)}_{q\leftarrow q}(\tauhat )C_F^{-1}
\right.\\[5mm]
&~~~\left. \dis
+                                                                                                                                                                  4\left(1+\tauhat^2\right)\left(\f{\ln (1-\tauhat )}{1-\tauhat}\right)_+
-2\f{1+\tauhat ^2}{1-\tauhat }\ln\tauhat +2(1-\tauhat ) \right],
\\[5mm]
\dis P^{(1)}_{q\leftarrow q}(\tauhat ) \hspace*{-2mm}
&=~\dis C_F\(\f{1+\tauhat^2}{1-\tauhat}\)_+ 
~=~C_F\[\f{1+\tauhat^2}{\(1-\tauhat\)_+}+\f{3}{2}\delta (1-\tauhat )\],
\ea
\label{eq:A1}
\ee
where the standard plus prescription $(\cdots )_+$ is given by
\be
\dis\int^1_0 d\alpha\, \xi (\alpha )\[\chi (\alpha )\]_+ \,=\,
\int_0^1 d\alpha\,\chi (\alpha )\[\xi(\alpha )-\xi (1)\].
\label{eq:plus}
\ee
In (\ref{eq:A1}),
the infrared $\dis\f{1}{\epsilon^2}$ poles cancel between 
$\Delta\widehat{\sigma}^{\rm virtual}_{\rm loop}$ and 
$\Delta\widehat{\sigma}^{\rm real}_{c\bbar}$.
The term $\Delta\widehat{\sigma}^{\rm virtual}_{\rm loop}$ 
from the virtual loop actually
contains two types of $\dis\f{1}{\epsilon}$ poles inside $[\cdots ]$ : 
$\dis\f{3}{\epsilon_{UV}}+\f{3}{\epsilon_{IR}}~$ 
 with $\epsilon_{UV}
=-\epsilon_{IR}\equiv \epsilon =(4-D)/2 >0$. Also, the 
$-\dis\f{3}{\epsilon}$ pole inside the Yukawa counter-term contribution
$\Delta\widehat{\sigma}^{\rm virtual}_{\rm count}$ is ultraviolet 
while the $+\dis\f{3}{\epsilon}$ pole inside
$\Delta\widehat{\sigma}^{\rm real}_{c\bbar}$ is infrared (IR). 
We see that the contribution 
$\Delta\widehat{\sigma}^{\rm virtual}_{\rm count}$ 
from the counter term of the Yukawa coupling is
crucial for cancelling the UV divergence from 
$\Delta\widehat{\sigma}^{\rm virtual}_{\rm loop}$ (which is absent in
the usual Drell-Yan type processes), while  
the soft $\dis\f{1}{\epsilon}$ divergences between
$\Delta\widehat{\sigma}^{\rm virtual}_{\rm loop}$
and $\Delta\widehat{\sigma}^{\rm real}_{c\bbar}$ cancel. Finally, 
the $\dis\f{1}{\epsilon}$ collinear singularity inside 
$\Delta\widehat{\sigma}^{\rm real}_{c\bbar}$  will be absorbed into
the re-definition of the PDF via the quark-quark transition function 
$P^{(1)}_{q\leftarrow q}(\tauhat )$. All the finite terms are
summarized in Eq.~(\ref{eq:NLO}).

\vspace*{1.5mm}
\noindent
{\bf B. Partonic processes $gc,g\bbar\to\phi^+X$}

The spin- and color-averaged amplitude-square for the 
$gc,g\bbar\to\phi^+X$ process is
\be
\overline{|{\cal M}|^2}=\dis
\f{\pi \alpha_s}{3(1-\epsilon )}\left(|\CL|^2+|\CR|^2\right)\mu^{2\epsilon}
\[(1-\epsilon )\(\f{\widehat{s}}{-\widehat{t}} 
   +\f{-\widehat{t}}{\widehat{s}} -2\)
   -2\f{\widehat{u}\,m_\phi^2}{\widehat{s}\,\widehat{t}}\].
\label{eq:M2gq}
\ee

The $O(\alpha_s)$ partonic cross section for 
the quark-gluon fusions is given by:
\be
\ba{ll}
\Delta\widehat{\sigma}^{\rm real}_{cg,\bar{b}g} \hspace*{-2mm}
& =\,
\dis\widehat{\sigma}_0\f{\alpha_sC_F}{2\pi}
\(\f{4\pi\mu^2}{Q^2}\)^{\epsilon}\[
\( -\f{1}{\epsilon}
 \f{\Gamma (1-\epsilon )}{\Gamma (1-2\epsilon )}
  +\ln\f{(1-\tauhat )^2}{\tauhat} \)
P^{(1)}_{q\leftarrow g}(\tauhat )+\f{1}{4}(-3+7\tauhat )(1-\tauhat )
 \],
\\[4.6mm]
P^{(1)}_{q\leftarrow g}(\tauhat ) \hspace*{-2mm}
& =\,\dis \f{1}{2}\[\tauhat^2+(1-\tauhat )^2 \],
\ea
\label{eq:A2}
\ee
where it is clear that the collinear $\dis\f{1}{\epsilon}$ singularity
will be absorbed into the re-definition of the PDF via the gluon-splitting
function $P^{(1)}_{q\leftarrow g}(\tauhat )$. The final result is finite
and is given in Eq.~(\ref{eq:NLO}).

\pagebreak

\addtolength{\topmargin}{0.6cm}

\end{document}
\end